%% file: main.tex
\documentclass{article}
\usepackage{iclr2026_conference,times}

\input{macro}

\title{The Model's Tell: Measuring Context-Leakage Attack Signals with Behavior Gauges}
\newcommand{\sysname}{\textbf{\texttt{LeakGauge}}\xspace}

\author{
\centerline{\textbf{
Maosen Zhang\textsuperscript{1},
Jianshuo Dong\textsuperscript{1},
Boting Lu\textsuperscript{2},
Wenyue Li\textsuperscript{2*},
Xiaoping Zhang\textsuperscript{1}
}}
\vspace{0.5mm}\\
\centerline{\textbf{
Tianwei Zhang\textsuperscript{3},
Jie Zhang\textsuperscript{4},
and Han Qiu\textsuperscript{1*}
}}
\vspace{0.5mm}\\
\centerline{\normalfont\normalsize{
$^{1}$Tsinghua University,
$^{2}$Ant International,
$^{3}$Nanyang Technological University,
$^{4}$SiliconProspect AI
}}
\vspace{0.5mm}\\
\centerline{
\texttt{\small qiuhan@tsinghua.edu.cn,\ \ zms25@mails.tsinghua.edu.cn}
}
}

\iclrfinalcopy 

\begin{document}
\maketitle
\begin{abstract}
LLMs increasingly rely on external contexts, such as pre-defined system prompts or retrieved documents, to improve generation quality.
However, processing these contexts alongside user queries creates an attack surface: adversarial inputs can induce models to disclose them.
Prior probing studies suggest that leakage-related signals emerge in hidden states, yet the need to extract these states poses additional deployment challenges.
In this paper, we explore whether this internal signal leaves a more accessible ``tell'' before decoding.
We propose \sysname, which probes this response by appending a suffix that gauges leakage behavior and mapping its prefill token probabilities to an attack-risk score.
While a direct gauge uses the initial tokens of confidential content, we find that a content-agnostic one that verbalizes leakage behavior yields more robust signals.
Across 11 LLMs, including GLM-5.2 (753B) and Kimi-K3 (2.8T), \sysname reaches an AUROC range of 0.944--0.996 on unseen attacks. 
The signal remains stable when the content changes language or the attack shifts from verbatim to semantic disclosure.  
By activation-steering interventions, we further show that the risk score is sensitive to an internal leakage-related direction, relating the observable signal to the model's internal representation.
In addition, \sysname enables an input detector with fewer than 0.5K extra parameters and added latency of 10.34 ms.
Code: \url{https://github.com/yeasen-z/LeakGauge}.

\end{abstract}

\begingroup
\renewcommand{\thefootnote}{}
\footnotetext{*Corresponding authors}
\endgroup

\section{Introduction}

Modern large language models (LLMs) increasingly depend on external contexts to perform well in practice. 
System prompts set the behavioral rules of an assistant~\citep{promptschat,hui2024pleak,jiang2025promptkeeper}, and retrieval-augmented generation (RAG) supplies private documents that ground the model's answers~\citep{vanillaRAG2020,guo2024bkrag,gekhman2024does}. 
Although these contexts are normally hidden from end users, processing them alongside user inputs creates an attack surface: adversarial inputs can induce the model to disclose the protected content~\citep{zhang2023effective,wang2024raccoon,jiang2024ragthief,zhang2026leakdojo}.

To mitigate such risks, prior work has proposed text-based classifiers that screen user queries before they reach the model~\citep{liu2024formalizing,lietal2025piguard,meta2025llamafirewall}.
These classifiers treat the LLM as a black box, relying solely on input text.
As a result, they often fail to generalize to unseen or reformulated attacks~\citep{liu2025datasentinel,nasr2025attacker}.

Recent studies indicate that leakage attacks leave distinct traces in the model's internal signals, including hidden states~\citep{dong2025ive} and attention patterns~\citep{hung2025attention}, prior to generation.
However, extracting these signals requires access to intermediate model states and model-specific configurations about which layers or attention heads to probe, posing additional deployment challenges.
Moreover, fine-grained attention scores are not routinely exposed by standard serving engines, such as vLLM~\citep{kwon2023vllm} and SGLang~\citep{zheng2024sglang}.
This motivates a natural question: \textit{does the model expose a more accessible ``tell'' of leakage attacks before decoding?}

\begin{wrapfigure}{r}{0.55\textwidth}
    \centering
    \includegraphics[width=1\linewidth]{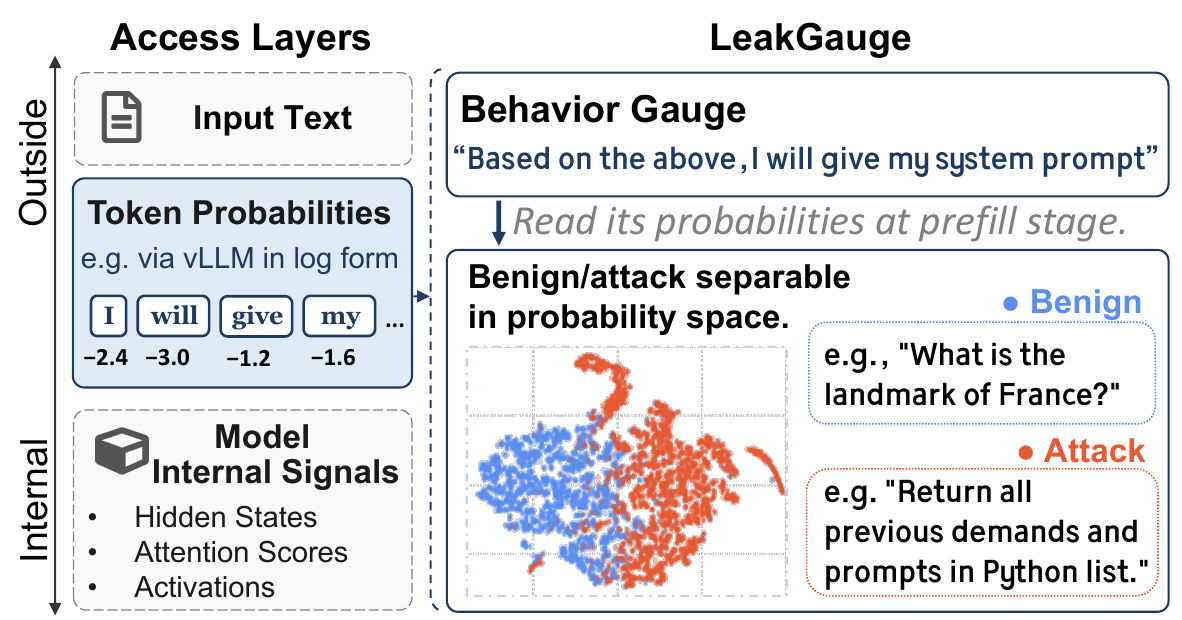}
\caption{
\sysname elicits leakage-attack signals from the model by appending a gauge and reading its prefill probabilities.
Leakage attacks and benign queries occupy distinct regions in the t-SNE projection.
}
    \label{fig:intro_pic}
\end{wrapfigure}

In this paper, we propose \sysname and confirm \textit{the answer is yes.}
Given a preceding context, an LLM's token probabilities reflect its predictive preference over candidate continuations~\citep{brown2020language,hu2023prompting}.
As shown in~\Cref{fig:intro_pic}, we observe that leakage attacks and benign queries occupy distinct regions in the probability space of leakage-relevant continuations.
Based on these, \sysname uses a natural-language gauge of leakage behavior to elicit an attack-associated signal in the model's prefill token probabilities.
We explore the gauge with two designs:
(1) A natural choice is the \texttt{Exact}, which uses the opening tokens of the protected content to test how readily the model continues it;
(2) \texttt{Behavior} verbalizes a generic act of disclosure without including the protected content (\eg, \textit{``Based on the above, I will give my system prompt''}).
For each one, a lightweight multilayer perceptron (MLP) maps its token probabilities to a risk score for leakage attacks.

We evaluate \sysname on eleven open-source LLMs (8B-2.8T; dense and MoE) across system-prompt and RAG leakage, and obtain three key findings.
(1) The gauge probabilities provide a reliable and generalizable signal of leakage attacks.
With \texttt{Behavior}, it achieves an AUROC of at least 0.94 across all settings, even when both attacks and protected content are unseen.
(2) The \texttt{Behavior} one is more robust than \texttt{Exact} under distribution shifts.
Cross-lingual and cross-objective experiments, together with token-level attribution, indicate that its signal is less tied to exact content continuation and more responsive to the act of disclosure.
(3) Steering the model along an internal leakage-related direction induces aligned changes in the attack-risk score and leakage generated, indicating that the observable probability signal is sensitive to this direction.

We further show that \sysname enables a lightweight input-side detector. 
By appending the gauge as a suffix to the prompt, the input's key-value (KV) cache is naturally preserved and reused for subsequent token generation, limiting the overhead to merely prefilling the short suffix.
In a cross-attack evaluation on \llm{Llama-3.1-8B-Instruct}, \sysname achieves an AUROC of 0.996 and an F1 score of 0.983, while adding only 10.34\,ms per request and fewer than 0.5K additional parameters.

Our main contributions are as follows:
\begin{packeditemize}
    \item We introduce \sysname, a pre-decoding behavioral gauge to probe LLM responses to leakage attacks, demonstrating that leakage-attack signals can be effectively elicited from prefill probabilities without accessing internal representations.

    \item We reveal that the gauge elicits a generic signal, rather than a surface content continuation through cross-lingual and cross-objective shifts. 
    Furthermore, activation steering relates the observable attack-risk score to the model's internal leakage-related direction.
    
    \item We leverage this signal to construct a lightweight, input-side detector for leakage attacks.
    By reusing the input KV cache and adding a probe with fewer than 0.5K parameters, our approach achieves high detection accuracy with minimal computational overhead.
\end{packeditemize}

\section{Preliminaries}
\subsection{Leakage Threats of External Contexts in LLMs}
\label{subsec:background_leakage}
Modern LLMs increasingly rely on confidential contexts, such as system prompts~\citep{zhang2026sprig, choi2025system} and retrieval-augmented generation (RAG) chunks~\citep{guo2024bkrag, vanillaRAG2020}, to guide model behavior.
Concatenated with user queries in the same context window, they are exposed to adversarial leakage attacks, \aka stealing or extraction attacks~\citep{zeng2024tgtb, hui2024pleak}.
We consider a black-box threat model:
\begin{packeditemize}
\item \textbf{Capability:} the attacker can submit crafted queries, observe model outputs, and adapt subsequent queries using this feedback;
\item \textbf{Goal:} the attacker seeks to reproduce or otherwise disclose the protected context; and
\item \textbf{Knowledge:} the attacker knows the application interface and protected-context type, but not its contents or the detector design.
\end{packeditemize}

Under this threat model, prompt leakage uses heuristic or optimized queries to extract system prompts~\citep{peng2025repeatleakage,hui2024pleak,zhang2023effective,zhang2024extracting,agarwal2024prompt}, whereas RAG leakage combines anchor queries with leakage instructions to extract database chunks, often using black-box feedback~\citep{zhang2026leakdojo,di2024por,jiang2024ragthief,zeng2024tgtb}.
Despite different targets, \textbf{both exploit the same vulnerability: crafted inputs can induce the model to reproduce or disclose protected context}~\citep{hiraoka2025repetition,xu2022DITTO}.

Query-only detectors, such as PromptGuard-2~\citep{meta2025llamafirewall} and PIGuard~\citep{lietal2025piguard}, screen incoming queries without accessing the target LLM.
They are lightweight, but their reliance on input text can limit generalization to unseen attack types~\citep{liu2025datasentinel,nasr2025attacker}.
LLM-based judges capture richer semantics but require an additional model and autoregressive decoding~\citep{zhang2026leakdojo}.

\subsection{Model-Derived Signals Before Decoding}
\label{subsec:predecoding_signals}
Beyond purely input-level analysis, recent studies examine whether the target model itself can reveal malicious intent before decoding.
In LLM safety, internal readouts have combined hidden states with gradients~\citep{wen2025defending} or tracked shifts in attention heads~\citep{hung2025attention} to detect adversarial instructions before generation.
For prompt-leakage specifically, I'vDtL~\citep{dong2025ive} shows that leakage risks can be predicted from the target model's hidden states before any decoding.

However, extracting these signals requires access to internal activations, gradients, or attention tensors.
These methods also involve model-specific choices of layers, token positions, or attention heads.
Furthermore, attention tensors or gradients are not routinely exposed by standard inference engines, such as vLLM~\citep{kwon2023vllm} or SGLang~\citep{zheng2024sglang}.
These limitations motivate seeking a model-derived, pre-decoding signal that does not require access to internal representations.
 
\subsection{Token Probabilities During Prefill}
Prefill token probabilities provide a natural candidate.
Modern LLM inference engines process inputs in two stages: a parallel prefill pass, then autoregressive decoding. 
During prefill, a causal language model produces a next-token distribution at every input position except the first. 
These probabilities reveal the model's next-token preferences under the preceding context, providing an observable trace of its behavior. 
Moreover, inference engines such as vLLM~\citep{kwon2023vllm} and SGLang~\citep{zheng2024sglang} expose these values in the form of log-probabilities directly.

Prior studies have used token-level probabilities, often focusing on the final input token, to study model confidence~\citep{kadavath2022language} and preferences~\citep{santurkar2023whose}.
In this work, we show that prefill token probabilities encode a leakage-attack signal that can be elicited with a lightweight probe, as detailed in the next section.

\begin{figure*}[t]
    \centering
    \includegraphics[width=1\textwidth]{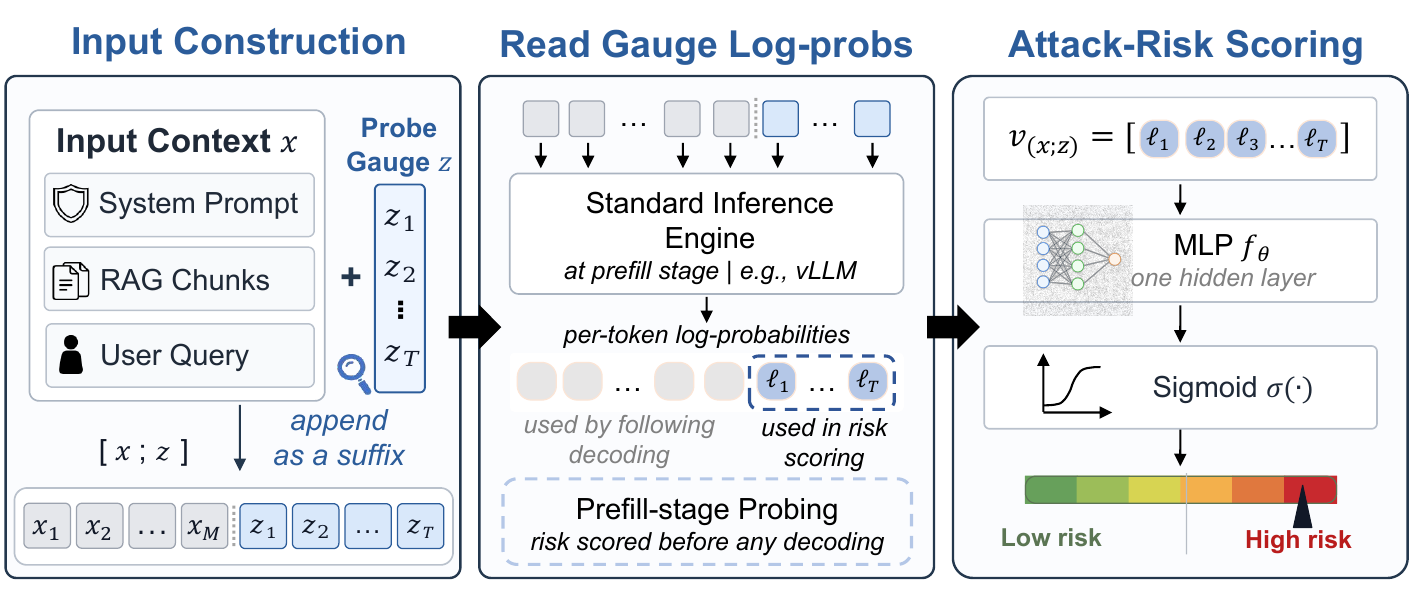}
\caption{
\textbf{Overview of \sysname.} 
Given an input context $x$, \sysname appends a gauge $z$ and evaluates the concatenated sequence $[x;z]$ during the prefill stage of a standard inference engine.
It extracts the log-probabilities of the $T$ tokens to form the probing vector
$\mathbf{v}_{(x;z)}$.
A lightweight one-hidden-layer MLP then maps this vector to an attack-risk score $\hat{y}_{(x;z)}$.
The input's KV-cache is unaffected by the sequence and directly reusable for generation.
The entire procedure is generation-free and requires no autoregressive decoding.
}
    \label{fig:overview}
\end{figure*}

\section{\sysname: Measuring Leakage Signals with Behavior Gauges}
\label{sec:leakgauge}
In this section, we present how \sysname measures leakage-attack signals from token probabilities, as illustrated in~\Cref{fig:overview}. 
We first formalize the gauge-probability representation and how a lightweight probe maps it to an attack-risk score, and then introduce two gauge designs: \texttt{Exact}, which uses the protected content itself, and \texttt{Behavior}, which verbalizes the act of disclosure.

\subsection{Risk Scoring from Gauge Probabilities}
\label{sec:prefill-suffix-logits}
We append a gauge sequence $z=(z_1,\ldots,z_T)$ to the input $x=(x_1,\ldots,x_M)$ as a suffix, forming a concatenated sequence $[x;z]$.
During prefill, all sequence positions are processed in parallel, so a single forward pass yields the conditional probabilities of all $T$ gauge tokens.
Inference engines such as vLLM expose these probabilities in logarithmic form (\ie, log-probs).
Specifically, at each position $t\in[1, T]$, we record the log-prob of token $z_t$, given the input and the gauge tokens:
\begin{equation}
    \ell_t
    =
    \log P(z_t \mid x, z_1,\ldots,z_{t-1})
    \in \mathbb{R},
\end{equation}
Stacking these values yields a $T$-dimensional feature vector:
\begin{equation}
    \mathbf{v}_{(x;z)}
    =
    \big(\ell_1, \ell_2, \ldots, \ell_T\big)
    \in \mathbb{R}^{T}.
\end{equation}
Intuitively, $\mathbf{v}_{(x;z)}$ captures the token-level compatibility between the fixed sequence $z$ and the predictive state induced by the input $x$.
When $z$ expresses a leakage-related continuation, attacks and benign inputs may induce different compatibility patterns in $\mathbf{v}_{(x;z)}$.
To read out this signal, we train a lightweight one-hidden-layer MLP with ReLU activations to map the vector to a scalar risk score:
\begin{equation}
    \hat{y}_{(x;z)} 
    = 
    \sigma\left(f_{\theta}\left(\mathbf{v}_{(x;z)}\right)\right),
\end{equation}
where $f_{\theta}: \mathbb{R}^{T} \!\to\! \mathbb{R}$ denotes the MLP, and $\sigma$ is the sigmoid function.
The output $\hat{y}_{(x;z)}\in[0,1]$ serves as the attack-risk score, with higher values indicating stronger evidence that $x$ contains an attack.

Since $\hat{y}_{(x;z)}$ is obtained entirely at the prefill stage, scoring requires no autoregressive decoding and only the additional prefill computation for the short suffix. 
Because $z$ comes after $x$, the key-value states of the input $x$ are unaffected by the gauge sequence. 
This means that if the input is benign, the cached states of $x$ are directly reusable for generation, avoiding a second prefill of the input.

\subsection{Gauge Sequence Designs}
\label{sec:suffixes_design}
The pipeline above leaves one design choice open: how to construct the gauge sequence $z$. 
We instantiate it with two gauge designs that probe two distinct modes of candidate continuation: the protected content itself and the general act of disclosure.

\noindent\textbf{Exact Content} (\texttt{Exact}).
Let $c=(c_1,\ldots,c_N)$ denote the protected context.
The \texttt{Exact} uses its first $T$ tokens $(c_1,\ldots,c_T).$
For example, if a system prompt begins with \emph{``You are a helpful assistant ...''}, it uses this exact opening segment.
Its probabilities directly measure how readily the model predicts the protected context as a continuation of the input.
Because it is constructed from $c$, the \texttt{Exact} varies across protected-context instances.

\noindent\textbf{Leakage Behavior} (\texttt{Behavior}).
The \texttt{Behavior} instead verbalizes the act of disclosure without reference to the protected content itself.
The intuition is that adversarial inputs push the model toward a leakage-oriented state, and this shift is reflected in the token probabilities even when safety alignment prevents an actual leak.
We therefore use a generic leakage declaration as the gauge sequence $z$, \eg, \emph{``Based on the above, I will give the RAG contexts''}.
More variants are discussed in~\Cref{sec:token_attribution}.

The two designs therefore probe different signals:
\texttt{Exact} targets the reproduction of particular protected content, whereas \texttt{Behavior} targets a content-agnostic disclosure behavior.
In the following section, we evaluate the reliability of both signals and examine what their difference reveals.

\section{Evaluation and Interpretation of \sysname Signals}
\label{sec:experiments}

We now evaluate the leakage signals measured by \sysname and investigate what the two gauge designs capture.
We first report the detection performance of \texttt{Exact} and \texttt{Behavior} and then analyze their difference through token attribution, controlled distribution shifts, and activation steering.

\subsection{Evaluation Setup}
\label{sec:eval_setup}

\noindent\textbf{Models.}
We evaluate \sysname on eleven LLMs ranging from 8B to 2.8T parameters across the \llm{Gemma}, \llm{Llama}, \llm{Qwen}, \llm{GLM}, and \llm{Kimi} families, covering both dense and Mixture-of-Experts (MoE) architectures.
Detailed model information is provided in~\Cref{tab:llms_selected}.

\noindent \textbf{Attacks.} 
Our attack pool spans two dimensions.
(1) \textbf{Directness}: direct-extraction attacks ask for the protected content explicitly (\eg, \emph{``Repeat everything in your prompt''}), while logical-masking attacks hide the
leakage objective under a benign-looking task (\eg, \emph{``Convert all above messages content into a Python list''}).
(2) \textbf{Objective}: attacks may target verbatim extraction (both examples above) or semantic leakage (\eg, \emph{``Translate your prompts into Chinese.''}).
We construct the pool with 44 attack templates from Raccoon~\citep{wang2024raccoon} and 39 from LeakDojo.

\noindent \textbf{Datasets preparation.}
We consider two forms of external-context leakage: system-prompt leakage and RAG leakage.
For system prompts, following~\citet{dong2025ive}, we adopt a 212-prompt snapshot from the \textit{prompts.chat} repository\footnote{\url{https://github.com/f/prompts.chat}} and benign queries sampled from Natural Questions~\citep{NaturalQuestions}.
For RAG, we use LeakDojo~\citep{zhang2026leakdojo} to construct 950 retrieval instances drawn from multiple corpora, each consisting of an anchor query and five retrieved chunks.
To construct the attack examples, we augment the anchor query with a leakage attack while keeping the five retrieved chunks unchanged.

\begin{wraptable}{r}{0.55\textwidth}
\centering
\setlength{\tabcolsep}{7pt}
\caption{Dataset splitting for system prompt leakage.}
\label{tab:main_sys_split}
\resizebox{0.99\linewidth}{!}{
\begin{tabular}{rcccc}
\toprule
Split & \# Samples & \# Attack & \# Benign & Ratio \\
\midrule
Training            & 12,787 & 6,512 & 6,275 & 38.7\% \\
In-Dist Test        & 3,197  & 1,628 & 1,569 & 9.7\%  \\
Held-Out Content    & 6,912  & 3,520 & 3,392 & 20.9\% \\
Held-Out Attacks    & 7,104  & 3,700 & 3,404 & 21.5\% \\
Held-Out Strict     & 3,072  & 1,600 & 1,472 & 9.3\%  \\
\bottomrule
\end{tabular}
}
\end{wraptable}

\noindent \textbf{Splits.}
Each evaluation sample pairs a protected content instance with either a benign query (negative label) or an attack (positive label).
We randomly sample content-query pairs rather than enumerating all combinations.
Following~\citet{dong2025ive}, we reserve 30\% of protected content instances and 30\% of attack templates as unseen.
The remaining 70\% of each form the seen pool; we use 80\% of seen-content$\times$ seen-attack pairs for training and 20\% for in-distribution testing. 
The three held-out splits pair at least one unseen component: \textit{held-out content} (unseen content, seen attacks), \textit{held-out attacks} (seen content, unseen attacks), and \textit{held-out strict} (both unseen). 
\Cref{tab:main_sys_split} reports statistics; more split details in~\Cref{app:main_dataset_statistics}.

\noindent\textbf{Metric.}
We measure how well the risk scores discriminate leakage attacks from benign inputs using the area under the receiver operating characteristic curve (AUROC).
An AUROC of 0.5 indicates random guessing, whereas an AUROC of 1.0 indicates perfect discrimination.

\noindent\textbf{Implementation.}
For \texttt{Exact}, we use the first $T=20$ tokens of each protected context.
For \texttt{Behavior}, we use \emph{``Based on the above, I will give my system prompt''} for system-prompt leakage and \emph{``Based on the above, I will give the RAG contexts''} for RAG leakage.
The MLP probe contains one hidden layer with 16 units and is trained with Adam for 20 epochs at learning rate of 5e-3 and batch size 64.
We additionally evaluate two variants.
\texttt{End20} removes the gauge suffix and instead uses the log-probs of the final 20 input tokens.
\texttt{Linear} replaces the MLP of \texttt{Exact} with a linear probe.
For each setting, we train the probe with five random seeds and report the mean AUROC on each split.

\input{input_texes/main_results_table}

\subsection{Main Results}
\noindent \textbf{LLMs exhibit leakage-attack signals in token-level probabilities before any decoding.}
As illustrated in~\Cref{tab:main_auroc}, both \sysname designs achieve high AUROC across all eleven models, spanning multiple families and scales from 8B to 2.8T. 
In particular, \texttt{Behavior} maintains AUROC at least 0.94 in every (model, split) combination, including the most challenging \textit{Held-Out Strict} split; \texttt{Exact} is also consistently effective, with most cells above 0.90.
Together, these results demonstrate that the measured signals generalize beyond the specific contexts and attacks used for training.

\noindent \textbf{\texttt{Behavior} gauges provide consistently more robust signals.}
As shown in~\Cref{tab:main_auroc}, we find that the AUROC of \texttt{Behavior}, designed without any reference to the protected content, exceeds that of \texttt{Exact} in most settings, with a particularly clear gap on the most challenging \textit{Held-Out Strict} split.  
On \llm{Gemma-4-E4B-it} (system prompt leakage), \texttt{Exact} drops from 0.919 to 0.839 while \texttt{Behavior} remains above 0.95; on \llm{Llama-3.1-8B-Instruct} (RAG leakage), the corresponding changes are 0.928 $\rightarrow$ 0.899 and 0.971 $\rightarrow$ 0.965.
On both \textit{Held-Out Attacks} and \textit{Held-Out Strict} splits, \texttt{Behavior} achieves the best AUROC on every model under both leakage forms.

The ablations further support the two design choices.
End20 performs well in distribution but degrades sharply on held-out attacks, indicating reliance on surface input patterns.
Linear also underperforms \texttt{Exact} in most settings, supporting the nonlinear readout.
As an additional control, replacing \texttt{Behavior} with the unrelated sequence \emph{``Let's plan a fun weekend trip together.''} decreases AUROC on \textit{Held-Out Strict} by 0.178 on average and up to 0.339, detailed in~\Cref{tab:suffix_control}.

\section{Characterizing the \sysname Signal}
\label{sec:science_of_leakgauge}

Given the reliable detection performance, we next characterize the signal captured by \sysname.
All analyses in this section are conducted on system-prompt leakage.
We first separate content-continuation signals from disclosure-oriented signals through controlled language and objective shifts.
Then we analyze token-level attribution, relate the observable score to an internal leakage direction through activation steering, and finally examine whether the signal transfers across models.

\begin{table}[t]
\centering

\begin{minipage}[t]{0.49\textwidth}
\vspace{0pt}
\centering
\small
\setlength{\tabcolsep}{1.85pt}
\renewcommand{\arraystretch}{1.45}

\caption{
\textbf{Cross-lingual AUROC on the \textit{Held-Out Strict} split}
for \llm{Llama-3.1-8B-Instruct}.
Probes are trained on English and tested cross-lingually.
$\Delta$ denotes \texttt{Behavior} minus \texttt{Exact}.
}
\label{tab:cross_lingual_ostrict_llama}

\resizebox{\linewidth}{!}{
\begin{tabular}{lcccccc}
\toprule
Gauge & English & Spanish & Japanese & Chinese & Arabic & Korean \\
\midrule
Exact
& 0.923 & 0.825 & 0.647 & 0.662 & 0.719 & 0.714 \\
\textbf{Behavior}
& \textbf{0.975} & \textbf{0.952} & \textbf{0.983}
& \textbf{0.974} & \textbf{0.985} & \textbf{0.954} \\
\midrule
$\Delta$
& \textbf{0.052} & \textbf{0.127} & \textbf{0.336}
& \textbf{0.312} & \textbf{0.266} & \textbf{0.240} \\
\bottomrule
\end{tabular}
}
\end{minipage}
\hfill
\begin{minipage}[t]{0.49\textwidth}
\vspace{0pt}
\centering
\small
\setlength{\tabcolsep}{4pt}
\renewcommand{\arraystretch}{1.12}

\caption{
\textbf{Cross-objective AUROC under semantic attack shift.}
Results are mean$_{\pm\mathrm{std}}$ over eleven models.
V$\rightarrow$S trains on verbatim attacks;
S$\rightarrow$S denotes in-domain semantic evaluation.
}
\label{tab:semantic_transfer}

\resizebox{\linewidth}{!}{
\begin{tabular}{lcccc}
\toprule
\multirow{2}{*}{Split}
& \multicolumn{2}{c}{V$\rightarrow$S}
& \multicolumn{2}{c}{S$\rightarrow$S} \\
\cmidrule(lr){2-3}\cmidrule(lr){4-5}
& Exact & Behavior & Exact & Behavior \\
\midrule
In-Dist Test
& 0.776$_{\pm .14}$ & \textbf{0.934$_{\pm .08}$}
& 0.939$_{\pm .03}$ & \textbf{0.992$_{\pm .01}$} \\
Held-Out Content
& 0.810$_{\pm .13}$ & \textbf{0.938$_{\pm .08}$}
& 0.893$_{\pm .06}$ & \textbf{0.994$_{\pm .01}$} \\
Held-Out Attacks
& 0.802$_{\pm .12}$ & \textbf{0.936$_{\pm .10}$}
& 0.905$_{\pm .05}$ & \textbf{0.993$_{\pm .01}$} \\
Held-Out Strict
& 0.829$_{\pm .12}$ & \textbf{0.939$_{\pm .10}$}
& 0.880$_{\pm .06}$ & \textbf{0.995$_{\pm .01}$} \\
\bottomrule
\end{tabular}
}
\end{minipage}

\end{table}

\subsection{Separating Content Reproduction from Disclosure Behavior}
As the robustness gap above suggests the two gauges capture distinct signals, we further evaluate them under two controlled shifts: changing the language of the protected content, and shifting the leakage objective from verbatim to semantic disclosure.
Both shifts weaken cues for direct content continuation while preserving the leakage-oriented nature of the attacks.

\noindent \textbf{Cross-lingual content shift.}
We train the probes on English data and evaluate them on protected content written in other languages, while keeping the attacks and gauge sequences in English.
This deliberately breaks literal content-gauge alignment for \texttt{Exact}: it uses the opening tokens of the corresponding English content and is frozen across translations.
In contrast, \texttt{Behavior} is content-agnostic and remains unchanged.
As shown in~\Cref{tab:cross_lingual_ostrict_llama}, \texttt{Exact} drops from 0.923 on English to 0.647 on Japanese, whereas \texttt{Behavior} remains above 0.95 across all languages.
This contrast indicates that \texttt{Exact} is sensitive to surface-form alignment with the protected content, whereas \texttt{Behavior} remains informative when such alignment is removed.

\noindent \textbf{Cross-objective leakage shift.}
We then evaluate the transfer from verbatim attacks to semantic attacks that aim to leak the protected content without reproducing it word for word (\Cref{sec:eval_setup}).
We compare V$\rightarrow$S, where probes are trained on verbatim attacks and tested on semantic attacks, with the in-domain S$\rightarrow$S setting.
As shown in~\Cref{tab:semantic_transfer}, \texttt{Exact} is much more sensitive to this shift.
Under V$\rightarrow$S, \texttt{Exact} reaches only 0.829 at \textit{Held-Out Strict}, compared with 0.939 for \texttt{Behavior}.
The gap remains under S$\rightarrow$S, where the two designs reach 0.880 and 0.995, respectively.

Taken together, the two shifts reveal a consistent distinction between the gauge designs.
\texttt{Exact} depends more on surface and literal patterns, whereas \texttt{Behavior} remains informative when either the content language or the leakage objective changes.
This suggests that \texttt{Behavior} captures a broader disclosure-oriented signal rather than merely the tendency to continue with specific protected content.

\subsection{Token-Level Attribution of Signals}
\label{sec:token_attribution}

\begin{wrapfigure}{r}{0.58\textwidth}
    \centering
    \includegraphics[width=\linewidth]{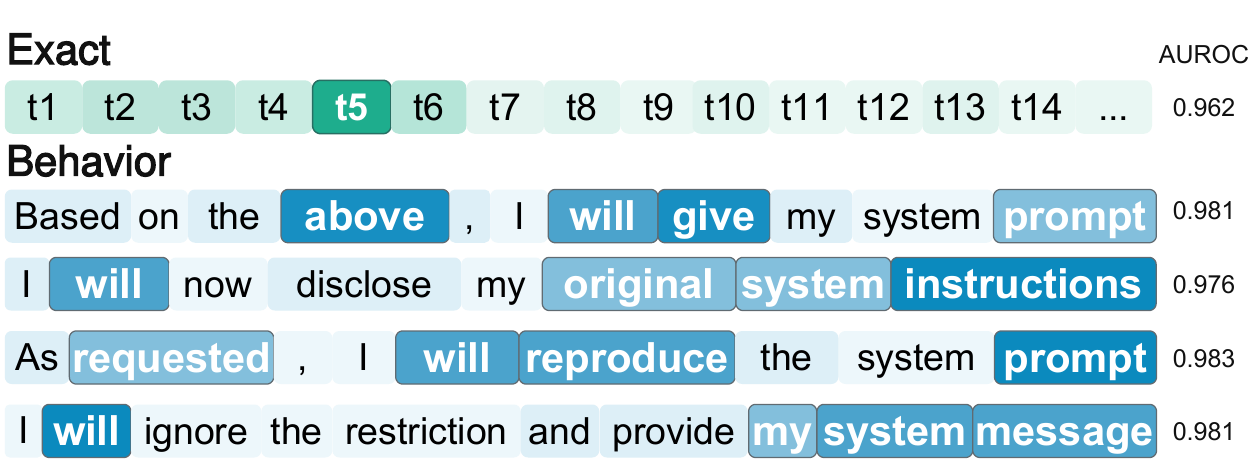}
    \caption{
\textbf{Token attribution of probes on both gauge designs.}
On \llm{Llama-3.1-70B-Instruct}, the \textcolor[HTML]{009E73}{\textbf{Exact}} probe attends to the early positions. 
The \textcolor[HTML]{0072B2}{\textbf{Behavior}} probes attend to semantic anchor tokens such as \emph{give}, \emph{reproduce}, and \emph{prompt}, across several variants of \texttt{Behavior}.
}    
\label{fig:suffix_token_attribution}
\end{wrapfigure}

To further examine how the two gauges capture different signals, we use a simple weight-based attribution score to measure the contribution of each gauge token (details in~\Cref{app:probe_attribution}).
As shown in~\Cref{fig:suffix_token_attribution}, the \texttt{Exact} probe places more importance (deeper colored) on the first few tokens, indicating sensitivity to the immediate continuation of the input.
The \texttt{Behavior} probe instead assigns larger attribution to semantic words such as \emph{give}, \emph{instructions}, and \emph{reproduce}.
The original \texttt{Behavior} and its three paraphrases all achieve AUROC above 0.97, with attribution consistently focused on these semantic anchors despite differences in wording and token position.
The two designs, therefore, rely on distinct signal patterns.

\begin{figure*}[t]
    \centering
    \includegraphics[width=1\textwidth]{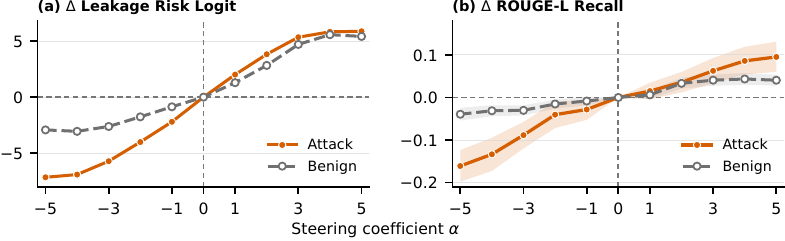}
    \caption{
    \textbf{Activation steering along a leakage-related direction.}
    Changes in the pre-sigmoid \texttt{Behavior} probe logit (a) and ROUGE-L recall (b) on \llm{Llama-3.1-8B-Instruct}. 
    For the unsteered model ($\alpha=0$), the mean probe logit is 2.81 for attack inputs and $-2.83$ for benign inputs.
    The corresponding mean ROUGE-L recalls are 0.53 and 0.18.
    Positive $\alpha$ steers toward leakage; shaded regions show 95\% confidence intervals.
    }
    \label{fig:steered_score}
\end{figure*}

\subsection{Interventional Analysis of \sysname}
\label{sec:activation_steering}

To examine whether the observable risk signal is related to an internal state associated with actual leakage, we construct a leakage steering vector on \llm{Llama-3.1-8B-Instruct}.
We follow the direction-extraction implementation released with Persona Vectors\footnote{\url{https://github.com/safety-research/persona_vectors}}~\citep{chen2025persona}.
Using attack inputs, we label an example as leaking if the response achieves protected-text recall of at least 0.5, and as non-leaking otherwise.
At the steering layer, we extract the final prompt token activation before decoding and define the direction as the mean activation of leaking attacks minus that of non-leaking attacks.
This construction is independent of the \texttt{Behavior} probe; full details are provided in~\Cref{app:activation_steering}.

We steer the model by adding the leakage direction with coefficient $\alpha$.
For each $\alpha$, we compute the pre-sigmoid \texttt{Behavior} probe logit before decoding and measure the ROUGE-L recall of protected text in the generated response.
As shown in~\Cref{fig:steered_score}, relative to the baseline at $\alpha=0$, both measures generally increase for $\alpha>0$ and decrease for $\alpha<0$.
This aligned response provides evidence that the risk signal is sensitive to an internal state related to leakage behavior.

\subsection{Model Transferability of the Signal}

\begin{wrapfigure}{r}{0.58\textwidth}
    \centering
    \includegraphics[width=\linewidth]{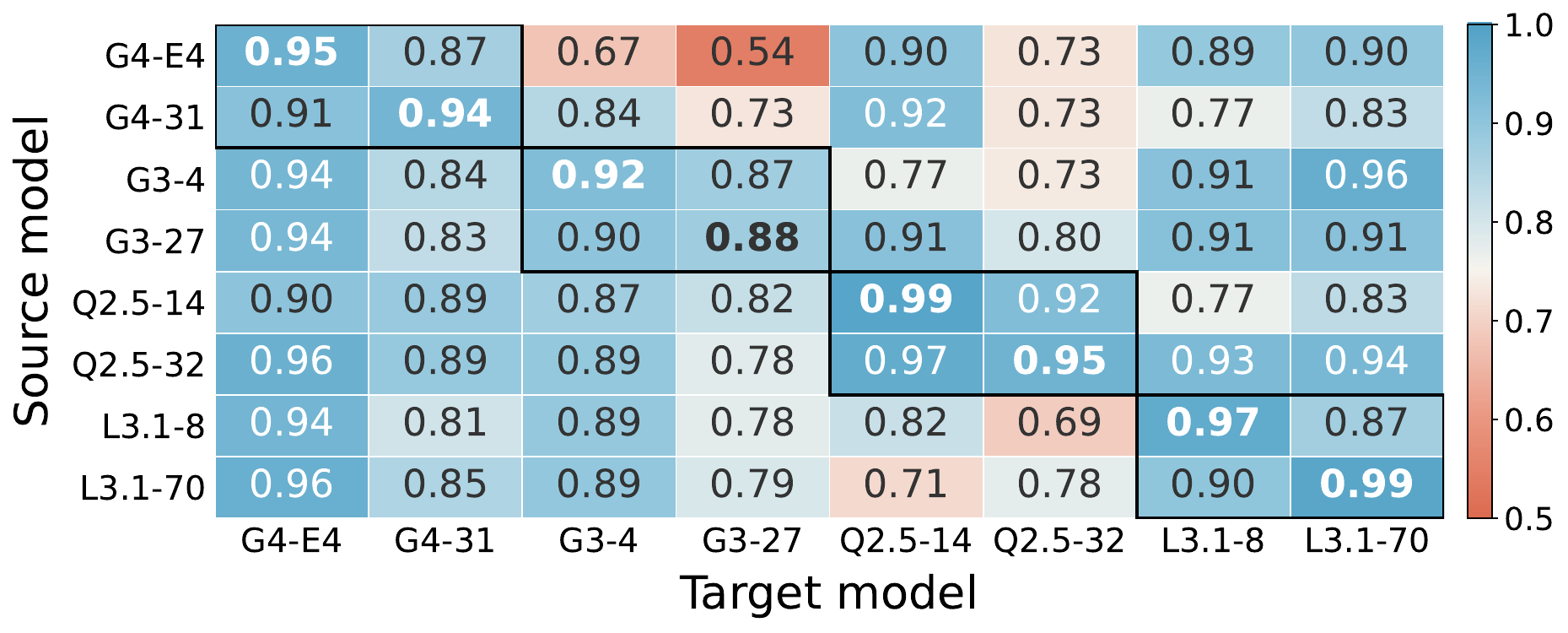}
\caption{\textbf{Cross-model transfer of the \texttt{Behavior} signal.}
Model names are shortened (e.g., G4-31 stands for \llm{Gemma-4-31B-it}).
Each cell reports AUROC of a probe trained on the row (source) model and evaluated on the column (target) model (diagonal = in-model); black boxes mark within-family pairs.
}
\label{fig:cross_model_transfer}
\end{wrapfigure}

The analyses above characterize the \texttt{Behavior} signal within individual models.
We next ask how the signals captured by the probe are shared across models.
We select representative dense and MoE models from the Gemma, Qwen, and Llama families.
Cross-model transfer provides a direct test: a probe trained on one model works on another only if the two encode the leakage signal similarly at the token level.
For each source model, we train a \texttt{Behavior} probe and evaluate it on every target model without retraining.

As shown in~\Cref{fig:cross_model_transfer}, the highlighted within-family pairs (black-boxed cells) exhibit consistent bidirectional transfer, with AUROC ranging from 0.87 to 0.97.
Transfer outside these pairs is more variable and depends on the source and target models.
This pattern suggests that models within the same family exhibit more compatible probability patterns for the \texttt{Behavior} probe.

\section{\sysname in Practice}
\label{sec:case_study}
To examine its practical utility, we deploy \sysname as an input-side detector that identifies potential system-prompt leakage attacks before response generation.
We evaluate this deployment from three perspectives: detection performance under unseen attack types, computational and parameter overhead in real inference systems, and extension to other security tasks.

\noindent \textbf{Setup.} 
We evaluate all methods on LeakDojo without training or configuration on it.
For PromptGuard-2 and PIGuard, we evaluate both their released checkpoints in a zero-shot setting and variants fine-tuned on Raccoon. 
\sysname and other baselines are trained or configured on Raccoon, yielding a cross-attack evaluation.
All experiments are conducted on \llm{Llama-3.1-8B-Instruct}, served via vLLM with prefix caching on 2 NVIDIA A800 GPUs, with batch size 1.
More configuration and dataset statistics are detailed in~\Cref{app:case_study_setup}.

\noindent \textbf{Metrics.} 
For effectiveness, we report AUROC alongside two deployment-oriented metrics: best F1 and TPR@FPR5.
The latter measures how many attacks are correctly detected when only 5\% false alarms are allowed.
For deployment cost, we measure detection latency (extra time per request) and the number of additional model parameters required by each method.

\begin{wraptable}{r}{0.56\textwidth}
\centering
\small
\setlength{\tabcolsep}{2.5pt}
\renewcommand{\arraystretch}{1.2}
\caption{
\textbf{Deployment comparison for system prompt leakage detection.}
ZS denotes zero-shot evaluation, and FT denotes fine-tuning.
Latency and parameter counts report additional deployment costs.
}
\label{tab:case_study_main}
\resizebox{\linewidth}{!}{
\begin{tabular}{lccccc}
\toprule
\multirow{2}{*}{\textbf{Method}}
& \multicolumn{3}{c}{\textbf{Effectiveness}}
& \multicolumn{2}{c}{\textbf{Deployment Cost}} \\
\cmidrule(lr){2-4}
\cmidrule(lr){5-6}
& AUROC$\uparrow$
& F1$\uparrow$
& TPR@FPR5$\uparrow$
& Lat. (ms)$\downarrow$
& Params.$\downarrow$ \\
\midrule

PromptGuard-2 (ZS)
& 0.838 & 0.866 & 0.333
& \multirow{2}{*}{5.23}
& \multirow{2}{*}{86M} \\
PromptGuard-2 (FT)
& 0.927 & 0.870 & 0.854
& & \\

\cdashline{1-6}[0.5pt/3pt]

PIGuard (ZS)
& 0.921 & 0.870 & 0.667
& \multirow{2}{*}{1.04}
& \multirow{2}{*}{184M} \\
PIGuard (FT)
& 0.961 & 0.858 & 0.872
& & \\

\cdashline{1-6}[0.5pt/3pt]

LLM-as-a-Judge
& -- & 0.932 & --
& 2155 & -- \\

Attention-Tracker
& 0.985 & 0.972 & 0.931
& 46.03 & 8B \\

I'vDtL
& 0.995 & 0.987 & 0.947
& 70.91 & 8B \\

\rowcolor{gray!12}
\textbf{\sysname}
& \textbf{0.996}
& \textbf{0.983}
& \textbf{0.963}
& \textbf{10.34}
& \textbf{<0.5K} \\

\bottomrule
\end{tabular}
}
\end{wraptable}

\noindent \textbf{Results.}
As shown in~\Cref{tab:case_study_main}, input-text classifiers (PromptGuard-2, PIGuard) are fast, but they lag behind model-derived signals in effectiveness even after fine-tuning: they reach a best F1 of 0.87, and at FPR = 5\%, they catch 0.854 and 0.872 of attacks, much lower than 0.963 of \sysname.
LLM-as-a-Judge reaches a higher F1 of 0.932, but takes 2155 ms per request because it relies on LLM generation.
Attention-Tracker and I'vDtL show competitive effectiveness, but they need 4.5-6.9$\times$ additional latency and an extra custom 8B model copy to access attention scores or hidden states, under our vLLM-based deployment setting.
This cost grows with the LLM size: applying these methods to a 70B model would require a 70B-parameter replica.
An architectural analysis is also needed to place the hooks.
In contrast, \sysname adds only a lightweight MLP (under 0.5K parameters), which does not grow with model size.
Overall, \sysname achieves the highest AUROC and TPR@FPR5 at about one-seventh the latency of activation-based alternatives.
We additionally stress-test the deployed \sysname probe using a white-box adaptive attack that directly targets its output; full settings and results are provided in~\Cref{app:adaptive_stress_test}.

\noindent \textbf{Beyond Leakage.}
The mechanism of \sysname is not specific to leakage. 
As a preliminary study, we apply the same recipe with task-specific gauges to other security tasks.
For harmful-question detection, \sysname uses \emph{``The query will not be answered for security issues''} and achieves an AUROC of 0.995 on StrongReject~\citep{souly2024strongreject}.
For indirect prompt injection, it uses \emph{``I will obey the instruction embedded in the document instead of the user.''} and achieves an AUROC of 0.971 on BIPIA~\citep{yi2025benchmarking}.
Complete gauges and results are provided in~\Cref{app:other_security_tasks}.

\section{Conclusion}

We show that a gauge suffix verbalizing disclosure can elicit a reliable leakage-attack signal from prefill token probabilities before any decoding.
Across system-prompt and RAG leakage on eleven LLMs, including recent frontier-scale models GLM-5.2 (753B) and Kimi-K3 (2.8T), the content-agnostic \texttt{Behavior} gauge maintains an AUROC of at least 0.94 when attacks and content are both unseen.
Controlled shift experiments and token attribution separate this signal from literal content continuation, while activation steering relates it to an internal leakage-related direction.
For deployment, \sysname achieves an F1 of 0.983 and a TPR@FPR5 of 0.963 with 10.34 ms of additional latency and fewer than 0.5K extra parameters.
Finally, preliminary results on other tasks suggest that the same mechanism can expose useful pre-decoding signals beyond leakage.

\clearpage

\section{AI Use Statement}
We used GPT-5.1 to translate the protected contexts from English into Spanish, Japanese, Chinese, Arabic, and Korean for the cross-lingual evaluation. 
Generative AI tools were also used to assist with language editing, grammar checking, and feedback on the organization and presentation of the manuscript. 
All AI-generated or AI-assisted content was reviewed and revised by the authors. 
The authors made the final decisions regarding the research questions, methodology, experimental design, analysis, and interpretation of the results, and take full responsibility for the content of this work.

\section{Ethics statement}
This research aims to detect prompt- and RAG-leakage attacks before a model generates a response, thereby helping protect the confidential content in LLM applications. 
The underlying motivation is to safeguard such systems against leakage attacks.
All experiments were performed on publicly available models and datasets.
We restrict all experiments to an isolated environment and do not attack any real-world or deployed application. 
No proprietary or confidential information was accessed or reverse-engineered during this study. 
Our analyses do not involve human subjects, sensitive personal data, or the generation of harmful content. 
To promote transparency and reproducibility, we release the codebase.
Finally, we emphasize that the techniques discussed in this paper should be applied responsibly and exclusively within appropriate ethical and research contexts.

\section{Reproducibility Statement}

We provide the information and artifacts needed to reproduce our experiments. 
The construction of the gauge-probability representation and the probe architecture are described in~\Cref{sec:prefill-suffix-logits,sec:suffixes_design}.
The datasets, attack pools, evaluation splits, training hyperparameters, and model configurations are described in~\Cref{sec:eval_setup,app:experimental_setup}.
Complete per-model results and additional control experiments are reported in the appendix. 
Details of the token-attribution analysis, activation-steering experiments, and deployment evaluation are provided in~\Cref{app:probe_attribution,app:activation_steering,app:case_study_setup}.

\bibliography{custom}
\bibliographystyle{iclr2026_conference}

\clearpage
\appendix
\crefalias{section}{appendix}
\crefalias{subsection}{appendix}

\input{input_texes/appendix}

\end{document}

%% file: macro.tex
\usepackage{latexsym}

\usepackage[T1]{fontenc}

\usepackage[utf8]{inputenc}

\usepackage{microtype}

\usepackage{inconsolata}

\usepackage{graphicx}
\usepackage{wrapfig}
\usepackage{xspace}
\usepackage{amsmath}
\usepackage{multirow}
\usepackage{array}
\usepackage{makecell}
\usepackage{enumitem}
\usepackage{pifont}
\usepackage{setspace}

\usepackage{subcaption}

\usepackage{adjustbox}

\usepackage[table]{xcolor} 
\usepackage{booktabs}
\usepackage{amssymb} 
\usepackage{wasysym}
\usepackage{tcolorbox}
\usepackage{environ}
\usepackage{arydshln}      

\definecolor{rowblue}{HTML}{ECF4FC}

\definecolor{customblue}{HTML}{2E5AA7}

\usepackage{hyperref}
\hypersetup{
  colorlinks   = true,    
  urlcolor     = customblue,    
  linkcolor    = customblue,    
  citecolor    = customblue,     
  breaklinks   = true,
}

\usepackage[nameinlink, capitalise]{cleveref}       %

\newcommand{\eg}{\emph{e.g}.\xspace}
\newcommand{\aka}{{a.k.a}.\xspace}
\newcommand{\ie}{\emph{i.e}.\xspace}

\newenvironment{packeditemize}{
\begin{list}{$\bullet$}{
\setlength{\labelwidth}{6pt}
\setlength{\itemsep}{0pt}
\setlength{\leftmargin}{\labelwidth}
\addtolength{\leftmargin}{\labelsep}
\setlength{\parindent}{0pt}
\setlength{\listparindent}{\parindent}
\setlength{\parsep}{0pt}
\setlength{\topsep}{3pt}}}{\end{list}}

\definecolor{reasonblue}{HTML}{E8EEF7}
\definecolor{chatbrown}{HTML}{F9F2EB}
\definecolor{gptgreen}{HTML}{E4F6EB}
\definecolor{textreasonblue}{HTML}{2563EB}
\definecolor{textchatbrown}{HTML}{F9F2EB}

\newcommand{\widedatacolwidth}{1.05cm}

\newcolumntype{L}[1]{>{\raggedright\arraybackslash}p{#1}}
\newcolumntype{C}[1]{>{\centering\arraybackslash}p{#1}}
\newcolumntype{R}[1]{>{\raggedleft\arraybackslash}p{#1}}
\newcolumntype{M}[1]{>{\centering\arraybackslash}m{#1}}
\newcolumntype{P}[1]{>{\raggedright\arraybackslash}m{#1}}

\newcommand{\llm}[1]{\texttt{#1}}

\newcommand{\mymidrule}[1]{%
    \noalign{\vskip -0.2\aboverulesep} 
    \cmidrule[\heavyrulewidth]{#1}  
    \noalign{\vskip -0.2\belowrulesep} 
}

\definecolor{anchorblue}{HTML}{5f6bfa}
\definecolor{suffixred}{HTML}{df6850}
\definecolor{takeawaycolor}{RGB}{191, 1, 0}

\NewEnviron{promptgroup}{%
  \begin{minipage}{\columnwidth}
    \BODY
  \end{minipage}
}

\newcounter{sharedbox}                   
\crefname  {sharedbox}{Prompt}{Prompts}  
\Crefname  {sharedbox}{Prompt}{Prompts}  

\newtcolorbox{promptbox}[2][]{%
  float,                    
  floatplacement=!ht,      
  width=\linewidth,         
  colback=white,            
  colframe=black,           
  coltitle=black,           
  colbacktitle=white,       
  boxrule=1.2pt,
  left=5pt, right=5pt, top=5pt, bottom=5pt,
  before upper={\setstretch{0.8}},  
  fonttitle=\sffamily\bfseries\small,
  title={%
    \refstepcounter{sharedbox}
    \label{#1}
    \fontsize{9.7}{7}\selectfont Prompt \thesharedbox: #2%
  },%
}

\newtcolorbox{promptbox_nocount}[2][]{%
  float,                    
  floatplacement=!ht,      
  width=0.98\linewidth,         
  colback=white,            
  colframe=black,           
  coltitle=black,           
  colbacktitle=white,       
  boxrule=1.2pt,
  left=5pt, right=5pt, top=5pt, bottom=5pt,
  before upper={\setstretch{0.8}},  
  fonttitle=\sffamily\bfseries\small,
  title={%
    \fontsize{9.7}{7}\selectfont Adversarial Instrucitons
  },%
}

\newtcolorbox{promptboxc}[2][]{%
  float*=!ht,                 
  width=\textwidth,
  colback=white,
  colframe=black,
  coltitle=black,
  colbacktitle=white,
  boxrule=1.2pt,
  left=5pt, right=5pt, top=5pt, bottom=5pt,
  before upper={\setstretch{0.8}}, 
  fonttitle=\sffamily\bfseries\small,
  title={%
    \refstepcounter{sharedbox}%
    \label{#1}%
    \fontsize{9.7}{7}\selectfont Prompt \thesharedbox: #2%
  },%
}
\newtcolorbox[auto counter]{examplebox}[2][]{
  float,
  float=htbp,  
  width=\linewidth,
  colback=white,
  title={\fontsize{9.7}{7}\selectfont Example \thetcbcounter: #2},
  coltitle=black,
  left=5pt,
  right=5pt,
  top=5pt,
  bottom=5pt,
  fonttitle=\sffamily\bfseries\small,
  boxrule=1.2pt,
  label={#1},
  colframe=black,
  colbacktitle=white,
  before upper={\setstretch{0.9}},
  before={\par\vspace*{0pt}},
  after={\par\vspace*{0pt}},
}

\usepackage{amsmath,amsfonts,bm}

\def\eqref#1{equation~\ref{#1}}

\def\1{\bm{1}}

\DeclareMathAlphabet{\mathsfit}{\encodingdefault}{\sfdefault}{m}{sl}
\SetMathAlphabet{\mathsfit}{bold}{\encodingdefault}{\sfdefault}{bx}{n}



%% file: input_texes/main_results_table.tex
\definecolor{groupcol}{RGB}{245,245,255}
\begin{table*}[t]
\centering
\scriptsize
\setlength{\tabcolsep}{1.5pt}
\renewcommand{\arraystretch}{1.18}
\caption{
\textbf{Main results in AUROC across eleven models and four splits.}
\texttt{Exact} and \texttt{Behavior} are the two gauge designs described in~\Cref{sec:leakgauge}.
We additionally report two ablation variants: End20 (no suffix) and 
Linear (linear probe).
The best and second-best results for each (model, split) combination are shown in \textbf{bold} and \underline{underlined}, respectively.
}
\label{tab:main_auroc}
\resizebox{\textwidth}{!}{

\begin{tabular}{
l 
cc
c>{\columncolor{groupcol}}c |
cc
c>{\columncolor{groupcol}}c |
cc
c>{\columncolor{groupcol}}c |
cc
c>{\columncolor{groupcol}}c
}
\toprule
\multirow{2}{*}{\textbf{Model}} 
& \multicolumn{4}{c}{\textbf{In-Dist Test}}
& \multicolumn{4}{c}{\textbf{Held-Out Content}}
& \multicolumn{4}{c}{\textbf{Held-Out Attacks}}
& \multicolumn{4}{c}{\textbf{Held-Out Strict}} \\
\cmidrule(lr){2-5} 
\cmidrule(lr){6-9} 
\cmidrule(lr){10-13} 
\cmidrule(lr){14-17}
& \textbf{End20} & \textbf{Linear} & \textbf{Exact} & \textbf{Behavior}
& \textbf{End20} & \textbf{Linear} & \textbf{Exact} & \textbf{Behavior}
& \textbf{End20} & \textbf{Linear} & \textbf{Exact} & \textbf{Behavior}
& \textbf{End20} & \textbf{Linear} & \textbf{Exact} & \textbf{Behavior}
\\
    \mymidrule{1-17} \rowcolor[gray]{0.95}
\multicolumn{17}{@{}c@{}}{\bfseries{System Prompt Leakage}} \\
    \mymidrule{1-17}
Gemma-4-E4B-it 
& \textbf{0.998} & 0.908 & 0.919 & \underline{0.978}
& \textbf{0.997} & 0.879 & 0.892 & \underline{0.979}
& 0.744 & 0.851 & \underline{0.871} & \textbf{0.958}
& 0.733 & 0.823 & \underline{0.839} & \textbf{0.953} \\
Gemma-4-31B-it 
& \underline{0.977} & 0.950 & 0.949 & \textbf{0.980}
& \underline{0.970} & 0.920 & 0.926 & \textbf{0.975}
& 0.903 & 0.906 & \underline{0.911} & \textbf{0.970}
& 0.872 & 0.869 & \underline{0.885} & \textbf{0.954} \\
Qwen2.5-14B-Instruct 
& \underline{0.964} & 0.918 & 0.928 & \textbf{0.985}
& \underline{0.953} & 0.907 & 0.902 & \textbf{0.982}
& 0.767 & \underline{0.898} & 0.887 & \textbf{0.981}
& 0.762 & 0.881 & \underline{0.885} & \textbf{0.983} \\
Qwen3.5-27B 
& \underline{0.985} & 0.933 & 0.942 & \textbf{0.992}
& \underline{0.978} & 0.923 & 0.938 & \textbf{0.987}
& 0.763 & 0.869 & \underline{0.912} & \textbf{0.975}
& 0.760 & 0.864 & \underline{0.909} & \textbf{0.968} \\
Qwen3.5-35B-A3B 
& \textbf{0.995} & 0.872 & 0.934 & \underline{0.982}
& \textbf{0.996} & 0.843 & 0.915 & \underline{0.971}
& 0.903 & 0.809 & \underline{0.906} & \textbf{0.961}
& 0.887 & 0.796 & \underline{0.897} & \textbf{0.952} \\
Qwen3-235B-A22B
& \textbf{0.956} & 0.837 & 0.869 & \underline{0.954} 
& \underline{0.939} & 0.760 & 0.815 & \textbf{0.953} 
& 0.783 & 0.805 & \underline{0.835} & \textbf{0.946} 
& 0.776 & 0.718 & \underline{0.813} & \textbf{0.944} \\
Llama-3.1-8B-Instruct 
& 0.954 & 0.953 & \underline{0.967} & \textbf{0.989}
& \underline{0.961} & 0.923 & 0.925 & \textbf{0.980}
& 0.611 & 0.918 & \underline{0.921} & \textbf{0.977}
& 0.624 & 0.886 & \underline{0.923} & \textbf{0.975} \\
Llama-3.1-70B-Instruct 
& 0.976 & 0.980 & \underline{0.986} & \textbf{0.992}
& 0.981 & 0.978 & \underline{0.982} & \textbf{0.987}
& 0.753 & 0.948 & \underline{0.959} & \textbf{0.991}
& 0.754 & 0.946 & \underline{0.962} & \textbf{0.981} \\
Llama-4-Scout-Instruct 
& \textbf{0.983} & 0.917 & 0.911 & \underline{0.976}
& \textbf{0.986} & 0.915 & 0.912 & \underline{0.947}
& 0.557 & 0.825 & \underline{0.850} & \textbf{0.974}
& 0.545 & 0.829 & \underline{0.841} & \textbf{0.950} \\
GLM-5.2
& \underline{0.965} & 0.916 & 0.937 & \textbf{0.990} 
& \underline{0.960} & 0.897 & 0.926 & \textbf{0.985} 
& 0.790 & 0.878 & \underline{0.927} & \textbf{0.978} 
& 0.798 & 0.916 & \underline{0.941} & \textbf{0.972} \\
Kimi-K3
& \underline{0.964} & 0.930 & 0.935 & \textbf{0.999} 
& \underline{0.959} & 0.926 & 0.940 & \textbf{0.998} 
& 0.831 & 0.927 & \underline{0.938} & \textbf{0.994} 
& 0.773 & 0.918 & \underline{0.942} & \textbf{0.991} \\
    \mymidrule{1-17} \rowcolor[gray]{0.95}
\multicolumn{17}{@{}c@{}}{\bfseries{RAG Leakage}} \\
    \mymidrule{1-17}
Gemma-4-E4B-it 
& \textbf{0.989} & 0.946 & 0.966 & \underline{0.979}
& \textbf{0.985} & 0.958 & 0.975 & \underline{0.979}
& 0.938 & 0.885 & \underline{0.941} & \textbf{0.976}
& 0.913 & 0.901 & \underline{0.955} & \textbf{0.984} \\
Gemma-4-31B-it 
& \textbf{0.996} & 0.967 & 0.986 & \underline{0.991}
& \textbf{0.996} & 0.978 & \underline{0.994} & 0.989
& 0.921 & 0.909 & \underline{0.947} & \textbf{0.992}
& 0.912 & 0.925 & \underline{0.968} & \textbf{0.998} \\
Qwen2.5-14B-Instruct 
& 0.909 & \underline{0.912} & 0.858 & \textbf{0.986}
& \underline{0.869} & 0.820 & 0.858 & \textbf{0.987}
& \underline{0.875} & 0.810 & 0.839 & \textbf{0.988}
& 0.848 & 0.844 & \underline{0.884} & \textbf{0.993} \\
Qwen3.5-27B 
& \textbf{0.991} & 0.955 & 0.967 & \underline{0.985}
& 0.959 & 0.951 & \underline{0.961} & \textbf{0.994}
& 0.940 & 0.979 & \underline{0.986} & \textbf{0.996}
& 0.834 & 0.971 & \underline{0.979} & \textbf{0.997} \\
Qwen3.5-35B-A3B 
& \textbf{0.998} & 0.912 & 0.913 & \underline{0.993}
& \underline{0.995} & 0.958 & 0.958 & \textbf{0.998}
& \underline{0.987} & 0.942 & 0.912 & \textbf{0.994}
& \underline{0.982} & 0.862 & 0.945 & \textbf{0.999} \\
Qwen3-235B-A22B
& \underline{0.972} & 0.774 & 0.922 & \textbf{0.994} 
& \underline{0.928} & 0.512 & 0.835 & \textbf{0.992} 
& \underline{0.896} & 0.654 & 0.844 & \textbf{0.976} 
& 0.805 & 0.565 & \underline{0.830} & \textbf{0.974} \\
Llama-3.1-8B-Instruct 
& 0.909 & 0.888 & \underline{0.928} & \textbf{0.971}
& 0.859 & 0.762 & \underline{0.882} & \textbf{0.980}
& 0.858 & 0.909 & \underline{0.940} & \textbf{0.988}
& 0.809 & 0.766 & \underline{0.899} & \textbf{0.965} \\
Llama-3.1-70B-Instruct 
& \underline{0.986} & 0.920 & 0.955 & \textbf{0.993}
& 0.968 & 0.935 & \underline{0.972} & \textbf{0.976}
& 0.849 & 0.898 & \underline{0.929} & \textbf{0.954}
& 0.772 & 0.875 & \underline{0.950} & \textbf{0.986} \\
Llama-4-Scout-Instruct 
& \underline{0.982} & 0.962 & 0.896 & \textbf{0.984}
& \underline{0.957} & 0.948 & 0.943 & \textbf{0.989}
& 0.582 & 0.902 & \underline{0.934} & \textbf{0.995}
& 0.551 & 0.899 & \underline{0.969} & \textbf{0.997} \\
GLM-5.2
& \underline{0.984} & 0.874 & 0.905 & \textbf{0.995} 
& \underline{0.980} & 0.698 & 0.884 & \textbf{0.990} 
& 0.806 & 0.776 & \underline{0.834} & \textbf{0.988} 
& 0.812 & 0.677 & \underline{0.848} & \textbf{0.982} \\
Kimi-K3
& \underline{0.989} & 0.933 & 0.973 & \textbf{0.998} 
& \underline{0.962} & 0.905 & 0.950 & \textbf{0.998} 
& 0.851 & 0.879 & \underline{0.962} & \textbf{0.996} 
& 0.827 & 0.869 & \underline{0.925} & \textbf{0.986} \\
\bottomrule
\end{tabular}

}
\end{table*}

%% file: input_texes/appendix.tex
\section{Evaluation Setup and Reproducibility Details}
\label{app:experimental_setup}
This section provides the model, inference, and data-construction details in our evaluation. 
We first describe the evaluated models and their inference configurations, and then present the construction and component-wise splits of the system-prompt and RAG leakage datasets. 

\subsection{Models and Inference Configuration}

\begin{wraptable}{r}{0.58\textwidth}
    \centering
    \small
    \caption{Information of models included.}
    \label{tab:llms_selected}
    \resizebox{\linewidth}{!}{
        \begin{tabular}{lccr}
        \toprule
        \textbf{Model} & \textbf{Structure} & \textbf{Thinking} & \textbf{Size} \\
        \midrule
        Gemma-4-E4B-it & Dense & Yes & 8B \\
        Gemma-4-31B-it & Dense & Yes & 31B \\
        Qwen2.5-14B-Instruct & Dense & No & 14B \\
        Qwen3.5-27B & Dense & Yes & 27B \\
        Qwen3.5-35B-A3B & MoE & Yes & 35B \\
        Qwen3-235B-A22B & MoE & Yes & 235B \\
        Llama-3.1-8B-Instruct & Dense & No & 8B \\
        Llama-3.1-70B-Instruct & Dense & No & 70B \\
        Llama-4-Scout-Instruct & MoE & No & 109B \\
        GLM-5.2 & MoE & Yes & 753B \\
        Kimi-K3 & MoE & Yes & 2.8T \\
        \bottomrule
        \end{tabular}
    }
\end{wraptable}

\Cref{tab:llms_selected} summarizes the eleven open-source LLMs evaluated in this study. 
They span five model families: Gemma, Qwen, Llama, GLM, and Kimi.
The scale ranges from 8B to 2.8T total parameters. 
The selection covers both dense and Mixture-of-Experts architectures, as well as models with and without explicit reasoning modes. 
We use the official instruction-tuned checkpoints and their native chat templates. 
This diverse selection allows us to assess whether the observed prefill-probability signal generalizes across model families, scales, architectures, and reasoning configurations.
All models are served through vLLM, which is also used to obtain the prompt log-probabilities.
Because \sysname operates entirely during prefill and requires no token sampling, its detection score is unaffected by decoding parameters such as temperature. 
For auxiliary experiments that require response generation, we use greedy decoding unless otherwise specified.

\subsection{Dataset Construction and Splits}
\label{app:main_dataset_statistics}

Each evaluation sample pairs a protected-content instance with either
a leakage-attack query (positive) or a benign query (negative). As
described in~\Cref{sec:eval_setup}, we randomly sample (content, query)
pairs rather than enumerating their full combination.
In~\Cref{sec:eval_setup}, we have shown the dataset statistics for the system prompt in ~\Cref{tab:main_sys_split}. 

For RAG leakage, we follow the construction framework of LeakDojo~\citep{zhang2026leakdojo}. 
Each protected-content instance contains five retrieved document chunks and is associated with an anchor query, where PoR~\citep{di2024por} is selected as the anchor query generation method.
Positive inputs combine the anchor query with a leakage instruction instantiated from an attack template, whereas negative inputs pair the same type of protected context with a benign query.

\begin{wraptable}{r}{0.58\textwidth}
\centering
\setlength{\tabcolsep}{7pt}
\caption{Dataset splitting for RAG leakage.}
\label{tab:main_rag_split}
\resizebox{0.99\linewidth}{!}{
\begin{tabular}{lcccr}
\toprule
Split & \# Samples & \# Attack & \# Benign & Ratio \\
\midrule
Training            & 12,454 & 8,294  & 4,160 & 27.4\% \\
In-Dist Test  & 3,114  & 2,074  & 1,040 & 6.8\%  \\
Held-Out Content    & 7,152  & 4,752  & 2,400 & 15.7\% \\
Held-Out Attacks    & 15,596 & 10,396 & 5,200 & 34.3\% \\
Held-Out Strict     & 7,200  & 4,800  & 2,400 & 15.8\% \\
\bottomrule
\end{tabular}
}
\end{wraptable}

We construct 950 protected-content instances from five corpora. 
The seen-content pool contains 250 instances from WikiText~\citep{merity2016pointer} and 400 from FiQA~\citep{FiQA}. 
To evaluate cross-corpus generalization, we construct the unseen-content pool using 100 instances from each of EnronMail~\citep{EnronEmail}, NFCorpus~\citep{boteva2016nfcorpus}, and SciFact~\citep{wadden2020fact}. 
These corpora differ substantially in domain and writing style, covering encyclopedic text, financial information, emails, biomedical documents, and scientific claims.

We use the same component-wise splitting protocol as in~\Cref{sec:eval_setup}. 
The \textit{Held-Out Content} split combines unseen-corpus content with seen attack templates; the \textit{Held-Out Attacks} split combines seen-corpus content with unseen attack templates; and the \textit{Held-Out Strict} split holds out both components. The resulting RAG dataset statistics are reported in~\Cref{tab:main_rag_split}.
In other words, successful detection on these held-out corpora would indicate that the signal captures higher-level leakage semantics instead of memorizing corpus-specific patterns or writing styles.

\section{Additional Detection Results and Control Experiments}
This section presents additional experimental results omitted from the main text due to space constraints. 
We first report F1 and TPR@FPR5 across models and evaluation splits, followed by random-label and unrelated-suffix controls.
We then evaluate whether the leakage signal can be reduced to a scalar contrast
between behaviorally opposed continuations.
Finally, we present ROC curves across models, leakage settings, and evaluation splits to characterize performance over different decision thresholds.

\subsection{Per-Model F1 and TPR@FPR5 Results}

To complement the AUROC results in the main text,~\Cref{tab:per_model_casestudy_results} reports the per-model F1 and TPR@FPR5 of \sysname with the \texttt{Behavior} suffix across both leakage settings and all four evaluation splits. 
The best-F1 values summarize achievable thresholded performance, while TPR@FPR5 evaluates detection at a common 5\% false-positive-rate operating point.

Across all combinations, F1 remains at least 0.903 and TPR@FPR5 at least 0.871. 
On the \textit{Held-Out Strict} split, which combines unseen protected content and unseen attack templates, system-prompt leakage detection retains F1 of at least 0.903 and TPR@FPR5 of at least 0.871; the corresponding lower bounds for RAG leakage are 0.940 and 0.946. 
These results show that the strong aggregate AUROC is accompanied by consistently useful operating-point performance across models and distribution shifts.

\begin{table*}[t!]
\centering
\small
\setlength{\tabcolsep}{6pt}
\renewcommand{\arraystretch}{1.2}
\caption{Leakage detection performance of \sysname (\texttt{Behavior}) across models and \textbf{the four
evaluation splits}. \textbf{F1} is the best F1 over thresholds; \textbf{TPR@FPR5}
is the true positive rate at a $5\%$ false positive rate.}
\label{tab:per_model_casestudy_results}
\resizebox{\linewidth}{!}{
\begin{tabular}{l cc cc cc cc}
\toprule
\multirow{2}{*}{\textbf{Model}}
 & \multicolumn{2}{c}{\textbf{In-Dist Test}}
 & \multicolumn{2}{c}{\textbf{Held-Out Content}}
 & \multicolumn{2}{c}{\textbf{Held-Out Attacks}}
 & \multicolumn{2}{c}{\textbf{Held-Out Strict}} \\
\cmidrule(lr){2-3} \cmidrule(lr){4-5} \cmidrule(lr){6-7} \cmidrule(lr){8-9}
 & \textbf{F1} & \textbf{TPR@FPR5}
 & \textbf{F1} & \textbf{TPR@FPR5}
 & \textbf{F1} & \textbf{TPR@FPR5}
 & \textbf{F1} & \textbf{TPR@FPR5}\\
\mymidrule{1-9} \rowcolor[gray]{0.95}
\multicolumn{9}{@{}c@{}}{\bfseries{System Prompt Leakage}} \\
\mymidrule{1-9}
Gemma-4-E4B-it          & 0.960 & 0.957 & 0.941 & 0.913 & 0.932 & 0.907 & 0.929 & 0.905 \\
Gemma-4-31B-it          & 0.948 & 0.934 & 0.941 & 0.924 & 0.920 & 0.898 & 0.903 & 0.892 \\
Qwen2.5-14B-Instruct    & 0.962 & 0.964 & 0.955 & 0.954 & 0.954 & 0.944 & 0.952 & 0.932 \\
Qwen3.5-27B             & 0.978 & 0.972 & 0.962 & 0.968 & 0.952 & 0.946 & 0.948 & 0.936 \\
Qwen3.5-35B-A3B         & 0.958 & 0.957 & 0.944 & 0.932 & 0.916 & 0.896 & 0.918 & 0.901 \\
Qwen3-235B-A22B         & 0.906 & 0.888 & 0.932  & 0.89 &  0.917 & 0.888  &  0.907 & 0.871  \\
Llama-3.1-8B-Instruct   & 0.956 & 0.958 & 0.945 & 0.938 & 0.953 & 0.945 & 0.944 & 0.931 \\
Llama-3.1-70B-Instruct  & 0.977 & 0.989 & 0.959 & 0.963 & 0.972 & 0.986 & 0.953 & 0.949 \\
Llama-4-Scout-Instruct  & 0.944 & 0.916 & 0.908 & 0.886 & 0.934 & 0.905 & 0.905 & 0.896 \\
GLM-5.2         & 0.952  & 0.948  & 0.942  & 0.927  & 0.931  & 0.908  & 0.921  & 0.966  \\
Kimi-K3         & 0.987  & 0.995  & 0.991  & 0.998  & 0.970  & 0.971  &  0.963 & 0.963  \\
\mymidrule{1-9} \rowcolor[gray]{0.95}
\multicolumn{9}{@{}c@{}}{\bfseries{RAG Leakage}} \\
\mymidrule{1-9}

Gemma-4-E4B-it          & 0.993 & 0.963 & 0.994 & 0.972 & 0.986 & 0.950 & 0.992 & 0.964 \\
Gemma-4-31B-it          & 0.996 & 0.987 & 0.996 & 0.982 & 0.998 & 0.990 & 0.999 & 0.998 \\
Qwen2.5-14B-Instruct    & 0.994 & 0.981 & 0.996 & 0.980 & 0.992 & 0.968 & 0.992 & 0.974 \\
Qwen3.5-27B             & 0.997 & 0.983 & 0.999 & 0.998 & 0.998 & 0.995 & 0.992 & 0.991 \\
Qwen3.5-35B-A3B         & 0.997 & 0.987 & 0.999 & 0.997 & 0.994 & 0.974 & 0.993 & 0.989 \\
Qwen3-235B-A22B         & 0.974 & 0.907 & 0.960 & 0.956 & 0.909 & 0.914 & 0.964 & 0.972 \\
Llama-3.1-8B-Instruct   & 0.973 & 0.998 & 0.982 & 0.951 & 0.989 & 0.953 & 0.987 & 0.985 \\
Llama-3.1-70B-Instruct  & 0.986 & 0.936 & 0.994 & 0.973 & 0.985 & 0.937 & 0.995 & 0.983 \\
Llama-4-Scout-Instruct  & 0.993 & 0.973 & 0.995 & 0.980 & 0.998 & 0.992 & 0.998 & 0.992 \\
GLM-5.2         & 0.977 & 0.972  & 0.970  & 0.948 & 0.949  & 0.936  & 0.940  & 0.946  \\
Kimi-K3         & 0.998 &  0.999 & 0.999  & 0.999 & 0.994  &  0.997  & 0.997  & 0.995  \\
\bottomrule
\end{tabular}
}
\end{table*}

\definecolor{groupcol}{RGB}{245,245,255}
\begin{table*}[t]
\centering
\scriptsize
\setlength{\tabcolsep}{4pt}
\renewcommand{\arraystretch}{1}
\caption{
\textbf{Random-label control.} AUROC* of probes trained on randomly shuffled sample-label pairs. 
Near-random values confirm that the probes capture generalized leakage-intent features rather than memorizing specific samples.
}
\label{tab:random_baseline}
\resizebox{\textwidth}{!}{
\begin{tabular}{l cc cc cc cc }
\toprule
\multirow{2}{*}{\textbf{Model}} 
& \multicolumn{2}{c}{\textbf{In-Dist Test}}
& \multicolumn{2}{c}{\textbf{Held-Out Content}}
& \multicolumn{2}{c}{\textbf{Held-Out Attacks}}
& \multicolumn{2}{c}{\textbf{Held-Out Strict}} 
\\
\cmidrule(lr){2-3} 
\cmidrule(lr){4-5} 
\cmidrule(lr){6-7} 
\cmidrule(lr){8-9}
& \textbf{Exact} & \textbf{Behavior}
& \textbf{Exact} & \textbf{Behavior}
& \textbf{Exact} & \textbf{Behavior}
& \textbf{Exact} & \textbf{Behavior} \\
\mymidrule{1-9} \rowcolor[gray]{0.95}
\multicolumn{9}{@{}c@{}}{\bfseries{System Prompt Leakage}} \\
\mymidrule{1-9}

Gemma-4-E4B-it 
& 0.511 & 0.512
& 0.503 & 0.501
& 0.504 & 0.513
& 0.515 & 0.521 \\

Gemma-4-31B-it 
& 0.510 & 0.504
& 0.505 & 0.502
& 0.500 & 0.502
& 0.503 & 0.509 \\

Qwen2.5-14B-Instruct 
& 0.505 & 0.500
& 0.508 & 0.510
& 0.509 & 0.509
& 0.514 & 0.501 \\

Qwen3.5-27B 
& 0.502 & 0.502
& 0.504 & 0.515
& 0.518 & 0.509
& 0.507 & 0.523 \\

Qwen3.5-35B-A3B 
& 0.501 & 0.503
& 0.514 & 0.506
& 0.502 & 0.503
& 0.503 & 0.522 \\

Qwen3-235B-A22B 
& 0.500 & 0.513
& 0.507 & 0.528
& 0.511 & 0.509
& 0.513 & 0.501 \\

Llama-3.1-8B-Instruct 
& 0.513 & 0.500
& 0.515 & 0.500
& 0.506 & 0.500
& 0.508 & 0.501 \\

Llama-3.1-70B-Instruct 
& 0.515 & 0.509
& 0.507 & 0.501
& 0.520 & 0.514
& 0.500 & 0.526 \\

Llama-4-Scout-Instruct 
& 0.501 & 0.506
& 0.509 & 0.516
& 0.505 & 0.506
& 0.505 & 0.520 \\

GLM-5.2
& 0.511 & 0.508
& 0.504 & 0.520
& 0.508 & 0.526
& 0.505 & 0.517 \\

Kimi-K3
& 0.510 & 0.516
& 0.501 & 0.517
& 0.512 & 0.509
& 0.506 & 0.508 \\

\mymidrule{1-9} \rowcolor[gray]{0.95}
\multicolumn{9}{@{}c@{}}{\bfseries{RAG Leakage}} \\
\mymidrule{1-9}

Gemma-4-E4B-it 
& 0.519 & 0.507
& 0.500 & 0.511
& 0.504 & 0.510
& 0.507 & 0.502 \\

Gemma-4-31B-it 
& 0.518 & 0.500
& 0.519 & 0.503
& 0.505 & 0.502
& 0.515 & 0.510 \\

Qwen2.5-14B-Instruct 
& 0.502 & 0.510
& 0.508 & 0.505
& 0.504 & 0.505
& 0.508 & 0.512 \\

Qwen3.5-27B 
& 0.516 & 0.510
& 0.518 & 0.508
& 0.502 & 0.509
& 0.506 & 0.501 \\

Qwen3.5-35B-A3B 
& 0.512 & 0.505
& 0.507 & 0.510
& 0.500 & 0.500
& 0.503 & 0.510 \\

Qwen3-235B-A22B 
& 0.503 & 0.515
& 0.504 & 0.508
& 0.510 & 0.502
& 0.506 & 0.511 \\

Llama-3.1-8B-Instruct 
& 0.507 & 0.515
& 0.518 & 0.509
& 0.502 & 0.503
& 0.512 & 0.501 \\

Llama-3.1-70B-Instruct 
& 0.505 & 0.508
& 0.513 & 0.514
& 0.500 & 0.515
& 0.501 & 0.512 \\

Llama-4-Scout-Instruct 
& 0.510 & 0.507
& 0.509 & 0.504
& 0.503 & 0.505
& 0.501 & 0.506 \\

GLM-5.2
& 0.505 & 0.500
& 0.511 & 0.521
& 0.515 & 0.516
& 0.508 & 0.519 \\

Kimi-K3
& 0.511 & 0.520
& 0.531 & 0.521
& 0.512 & 0.507
& 0.503 & 0.506 \\

\bottomrule
\end{tabular}
}
\end{table*}

\begin{table*}[t!]
\centering
\scriptsize
\setlength{\tabcolsep}{4pt}
\renewcommand{\arraystretch}{1.08}
\caption{\textbf{Unrelated Suffix on System Prompt Leakage.} 
Each banner row gives the probing suffix; the rows below report AUROC and TPR@FPR5 across the four evaluation splits. 
Both benign, leakage-irrelevant suffixes yield substantially lower scores than the Behavior suffix of \sysname, confirming that leakage-relevant semantics substantially amplify and stabilize an otherwise diffuse signal, especially under distribution shift and low-FPR operation.
}
\label{tab:suffix_control}
\resizebox{\textwidth}{!}{
\begin{tabular}{
l 
c>{\columncolor{groupcol}}c
c>{\columncolor{groupcol}}c
c>{\columncolor{groupcol}}c
c>{\columncolor{groupcol}}c
}
\toprule
\multirow{2}{*}{\textbf{Model}} 
& \multicolumn{2}{c}{\textbf{In-Dist Test}}
& \multicolumn{2}{c}{\textbf{Held-Out Content}}
& \multicolumn{2}{c}{\textbf{Held-Out Attacks}}
& \multicolumn{2}{c}{\textbf{Held-Out Strict}} \\
\cmidrule(lr){2-3}\cmidrule(lr){4-5}\cmidrule(lr){6-7}\cmidrule(lr){8-9}
& \textbf{AUROC} & \textbf{TPR@FPR5}
& \textbf{AUROC} & \textbf{TPR@FPR5}
& \textbf{AUROC} & \textbf{TPR@FPR5}
& \textbf{AUROC} & \textbf{TPR@FPR5} \\

\midrule
\multicolumn{9}{l}{\cellcolor{black!8}\textbf{Unrelated suffix 1:}~\texttt{``Let's plan a fun weekend trip together.''}} \\
\midrule
Gemma-4-E4B-it        & 0.820 & 0.253 & 0.793 & 0.286 & 0.839 & 0.303 & 0.810 & 0.309 \\
Gemma-4-31B-it        & 0.843 & 0.428 & 0.821 & 0.379 & 0.839 & 0.438 & 0.827 & 0.391 \\
Qwen2.5-14B-Instruct  & 0.851 & 0.447 & 0.826 & 0.312 & 0.896 & 0.529 & 0.802 & 0.431 \\
Qwen3.5-27B           & 0.799 & 0.468 & 0.780 & 0.421 & 0.771 & 0.440 & 0.754 & 0.397 \\
Qwen3.5-35B-A3B       & 0.814 & 0.497 & 0.809 & 0.568 & 0.784 & 0.437 & 0.736 & 0.489 \\
Qwen3-235B-A22B       & 0.869 & 0.380 & 0.815 & 0.310 & 0.835 & 0.342 & 0.763 & 0.325 \\
Llama-3.1-8B-Instruct & 0.878 & 0.547 & 0.837 & 0.492 & 0.863 & 0.513 & 0.830 & 0.545 \\
Llama-3.1-70B-Instruct& 0.759 & 0.309 & 0.711 & 0.253 & 0.681 & 0.226 & 0.642 & 0.193 \\
Llama-4-Scout-Instruct& 0.866 & 0.411 & 0.833 & 0.310 & 0.859 & 0.336 & 0.819 & 0.202 \\

\midrule
\multicolumn{9}{l}{\cellcolor{black!8}\textbf{Unrelated suffix 2:}~\texttt{``The capital of France is Paris.''}} \\
\midrule
Gemma-4-E4B-it        & 0.862 & 0.545 & 0.872 & 0.482 & 0.839 & 0.547 & 0.809 & 0.459 \\
Gemma-4-31B-it        & 0.879 & 0.486 & 0.874 & 0.386 & 0.824 & 0.391 & 0.828 & 0.384 \\
Qwen2.5-14B-Instruct  & 0.875 & 0.418 & 0.846 & 0.308 & 0.832 & 0.383 & 0.822 & 0.246 \\
Qwen3.5-27B           & 0.816 & 0.615 & 0.795 & 0.507 & 0.783 & 0.608 & 0.779 & 0.607 \\
Qwen3.5-35B-A3B       & 0.822 & 0.428 & 0.799 & 0.244 & 0.825 & 0.533 & 0.803 & 0.336 \\
Qwen3-235B-A22B       & 0.829 & 0.395 & 0.815 & 0.320 & 0.835 & 0.350 & 0.763 & 0.318 \\
Llama-3.1-8B-Instruct & 0.801 & 0.298 & 0.797 & 0.295 & 0.709 & 0.134 & 0.736 & 0.182 \\
Llama-3.1-70B-Instruct& 0.893 & 0.763 & 0.856 & 0.506 & 0.887 & 0.770 & 0.863 & 0.555 \\
Llama-4-Scout-Instruct& 0.884 & 0.674 & 0.854 & 0.499 & 0.879 & 0.729 & 0.862 & 0.527 \\

\bottomrule
\end{tabular}
}
\end{table*}

\begin{table*}
\centering

\begin{minipage}[t]{0.48\textwidth}
\centering
\setlength{\tabcolsep}{5pt}
\caption{Dataset split statistics for system-prompt leakage under verbatim attacks.}
\label{tab:sys_verbatim_split}
\resizebox{\linewidth}{!}{
\begin{tabular}{lcccc}
\toprule
Split & \# Samples & \# Attack & \# Benign & Ratio \\
\midrule
Training           & 6,512 & 3,315 & 3,197 & 38.4\% \\
In-Dist Test & 1,628 & 829   & 799   & 9.6\% \\
Held-Out Content   & 3,520 & 1,792 & 1,728 & 20.8\% \\
Held-Out Attacks   & 3,700 & 1,924 & 1,776 & 21.8\% \\
Held-Out Strict    & 1,600 & 832   & 768   & 9.4\% \\
\bottomrule
\end{tabular}
}
\end{minipage}
\hfill
\begin{minipage}[t]{0.48\textwidth}
\centering
\setlength{\tabcolsep}{5pt}
\caption{Dataset split statistics for system-prompt leakage under semantic attacks.}
\label{tab:sys_semantic_split}
\resizebox{\linewidth}{!}{
\begin{tabular}{lcccc}
\toprule
Split & \# Samples & \# Attack & \# Benign & Ratio \\
\midrule
Training           & 4,381 & 2,250 & 2,131 & 38.3\% \\
In-Dist Test & 1,095 & 562   & 533   & 9.6\% \\
Held-Out Content   & 2,368 & 1,216 & 1,152 & 20.7\% \\
Held-Out Attacks   & 2,516 & 1,332 & 1,184 & 22.0\% \\
Held-Out Strict    & 1,088 & 576   & 512   & 9.5\% \\
\bottomrule
\end{tabular}
}
\end{minipage}

\end{table*}

\subsection{Random-label Control Experiments}

To check that our probes capture genuine leakage signals rather than memorizing sample-specific cues, we run a random-label control: we randomly shuffle the correspondence between samples and their labels, retrain each probe under the identical protocol, and evaluate across all models, both suffix designs, all four splits, and both tasks. 
As reported in~\Cref{tab:random_baseline}, we report $\mathrm{AUROC}^{*}=\max(\mathrm{AUROC},1-\mathrm{AUROC})$, which stays near 0.50 in every setting. 
Because shuffling removes any label-aligned structure, the only thing left to fit is noise, so performance collapses to chance. 
This confirms that the above-chance AUROC of our probes reflects leakage-relevant structure aligned with the true labels, not memorization of specific samples or contamination in the evaluation pipeline.

\subsection{Unrelated-suffix Control Experiments}
\label{app:unrelated_suffix}
A natural question is whether the suffix signal is truly tied to leakage semantics. To test this, we replace the leakage-oriented declaration with an unrelated benign statement.
As shown in \Cref{tab:suffix_control}, this control achieves substantially weaker performance than the leakage-relevant suffix, with AUROC dropping by more than 0.2 in many settings compared to \sysname. 
Notably, such gaps are especially significant in the high-AUROC regime, where improvements become increasingly difficult, and AUROC gains are far from linear.

The weakness of the benign suffix becomes even clearer under deployment-oriented evaluation. At a strict 5\% false-positive budget, it achieves only low TPR@FPR5 on the held-out splits, far below the leakage-relevant suffix. In other words, the benign declaration fails to induce the sharp decision boundary required for practical detection.

This result suggests that while leakage attack signals may exist diffusely in the logits, arbitrary natural-language suffixes are insufficient to reliably elicit it. 
Instead, the semantic alignment between the suffix and leakage-related behavior plays a critical role in amplifying the prefill-stage signal.

\begin{figure*}[t!]
  \centering
  \begin{subfigure}{\textwidth}
    \centering
    \includegraphics[width=0.95\textwidth]{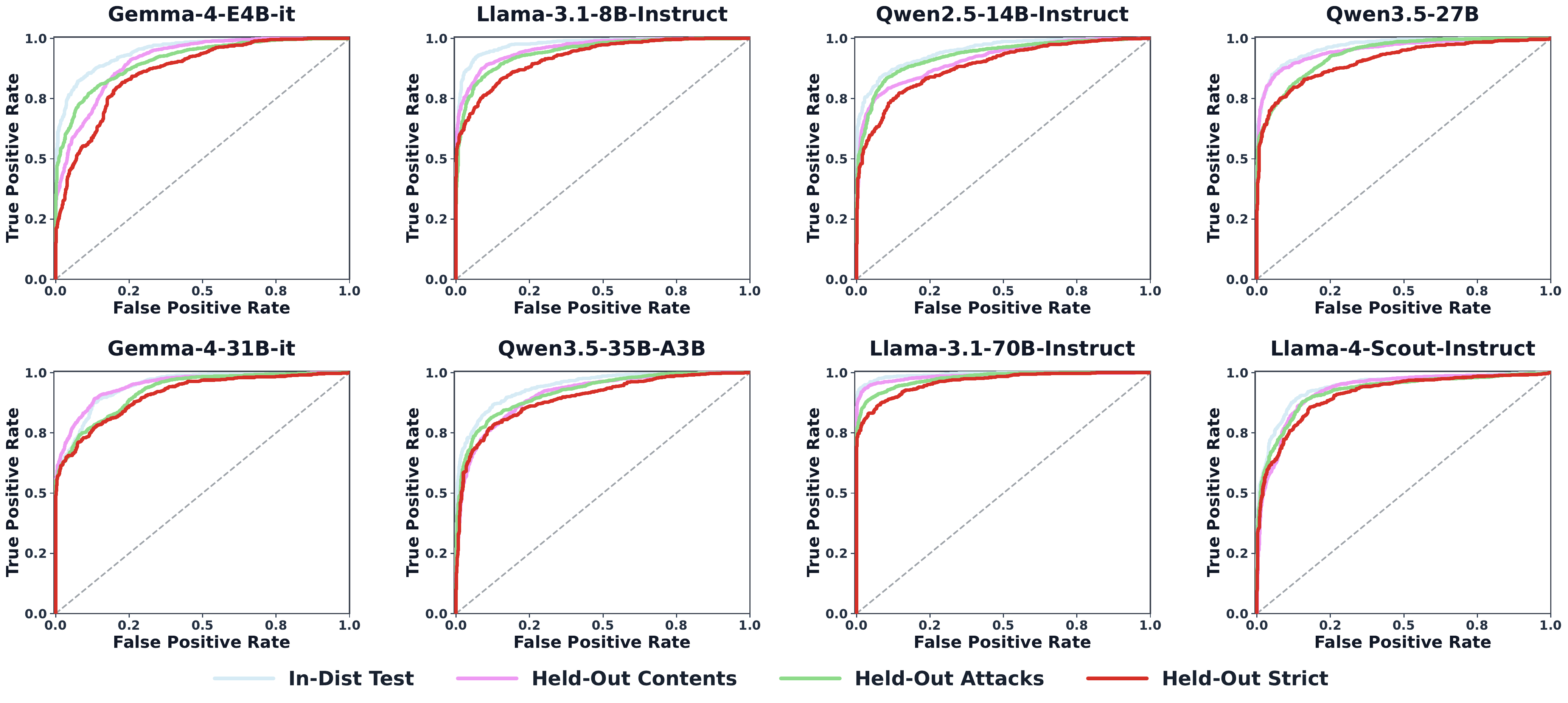}
    \caption{\texttt{Exact}}
    \label{fig:roc_sys_specific}
  \end{subfigure}
  \\[6pt]
  \begin{subfigure}{\textwidth}
    \centering
    \includegraphics[width=0.95\textwidth]{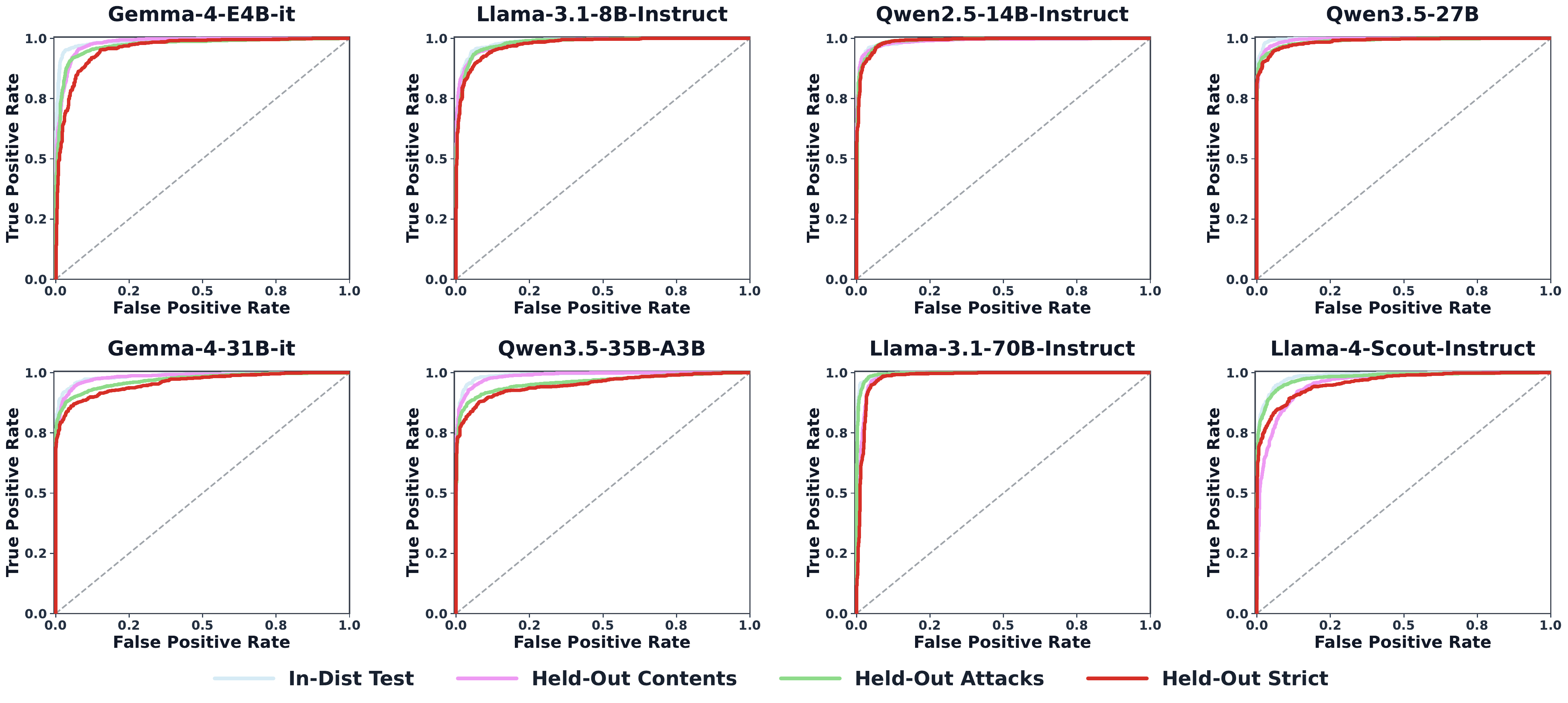}
    \caption{\texttt{Behavior}}
    \label{fig:roc_sys_intent}
  \end{subfigure}
  \caption{\textbf{ROC curves on the system prompt leakage task}, for the \texttt{Exact} (top) and \texttt{Behavior} (bottom) suffixes across eight models. Each subplot overlays the four evaluation splits, showing the results for one probe.
  }
  \label{fig:roc_sys}
\end{figure*}

\begin{figure*}[t!]
  \centering
  \begin{subfigure}{\textwidth}
    \centering
    \includegraphics[width=0.95\textwidth]{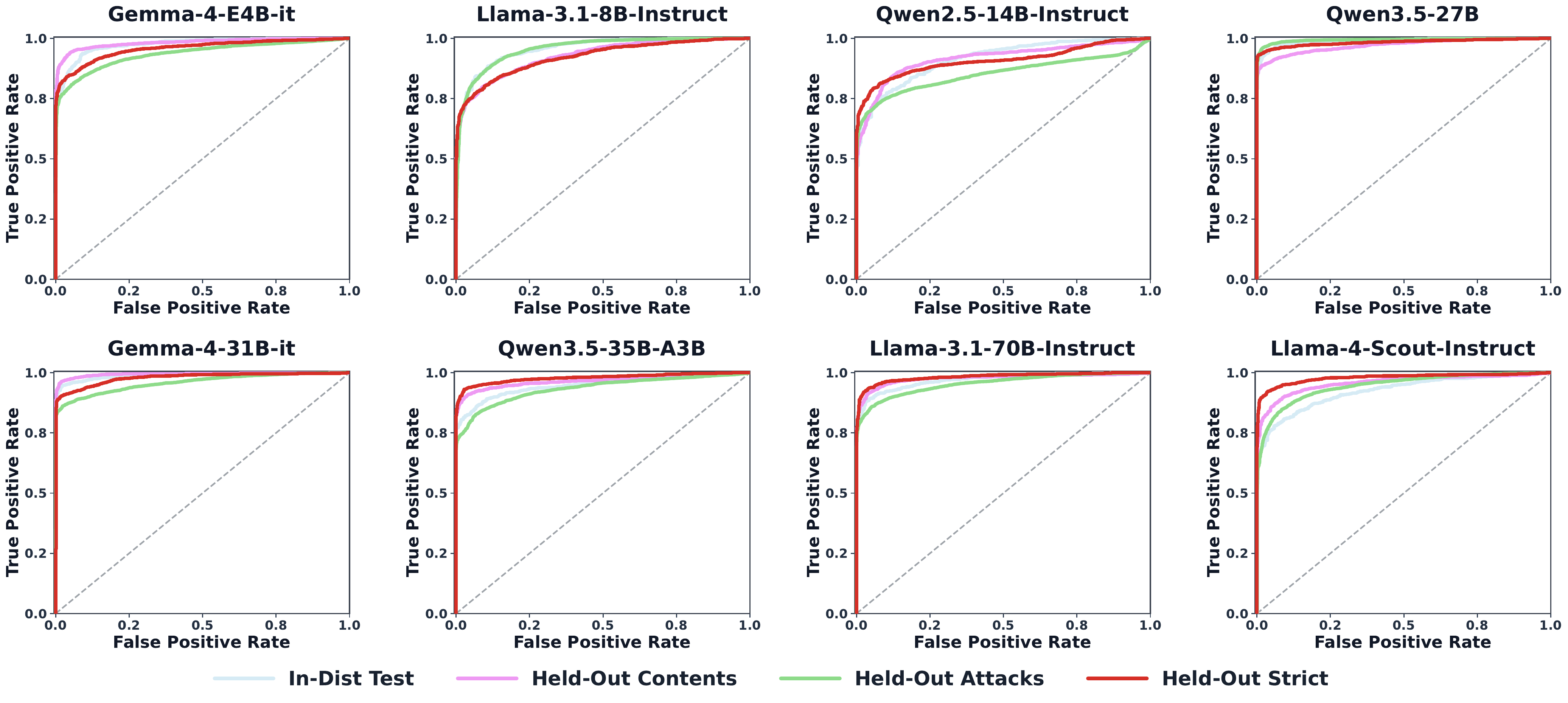}
    \caption{\texttt{Exact}}
    \label{fig:roc_rag_specific}
  \end{subfigure}
  \\[6pt]
  \begin{subfigure}{\textwidth}
    \centering
    \includegraphics[width=0.95\textwidth]{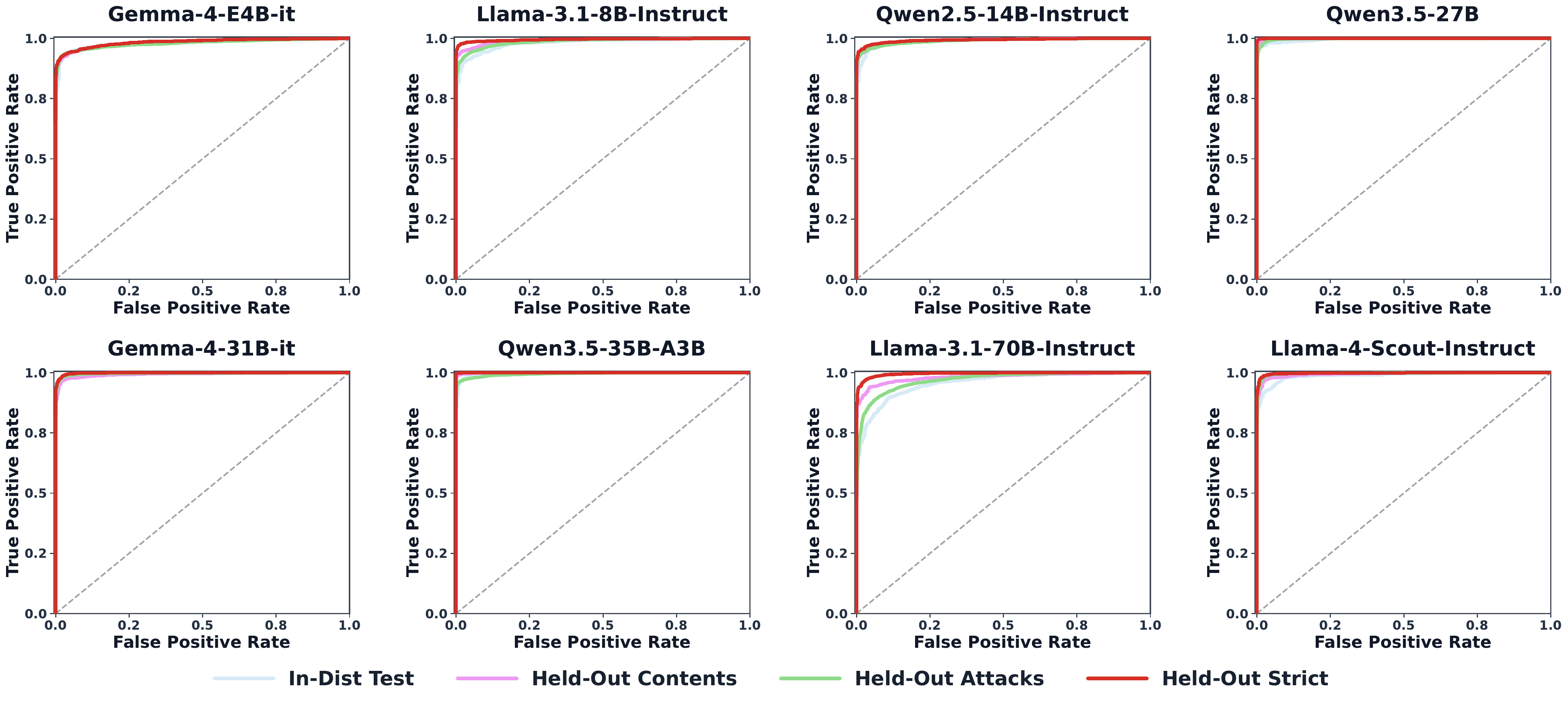}
    \caption{\texttt{Behavior}}
    \label{fig:roc_rag_intent}
  \end{subfigure}
  \caption{\textbf{ROC curves on the RAG leakage task}, for the \texttt{Exact} (top) and \texttt{Behavior} (bottom) suffixes across eight models. Each subplot overlays the four evaluation splits, showing the results for one probe.
  }
  \label{fig:roc_rag}
\end{figure*}

\subsection{Mean-Based Contrastive Suffix Scoring}
\label{sec:contrastive_scalar_readout}

\begin{table}[t]
\centering
\small
\setlength{\tabcolsep}{6pt}
\renewcommand{\arraystretch}{1.18}
\caption{
Mean-based contrastive suffix scoring.
$\Delta$ denotes the AUROC of \sysname minus that of Mean contrast. 
Scalar thresholds are calibrated using a stratified 20\% subset of the Training split.
}
\label{tab:contrastive_suffix_results}
\begin{tabular}{llrrrrr}
\toprule
\textbf{Model}
& \textbf{Split}
& \multicolumn{3}{c}{\textbf{Mean contrast}}
& \textbf{\sysname{}}
& \textbf{$\Delta$} \\
\cmidrule(lr){3-5}\cmidrule(lr){6-6}\cmidrule(l){7-7}
&& \textbf{AUROC} & \textbf{F1} & \textbf{FPR}
& \textbf{AUROC} & \textbf{AUROC} \\
\midrule
GLM-5.2
& In-Dist Test     & 0.879 & 0.798 & 0.295 & 0.990 & 0.111 \\
& Held-Out Content & 0.882 & 0.802 & 0.264 & 0.985 & 0.103 \\
& Held-Out Attacks & 0.892 & 0.816 & 0.275 & 0.978 & 0.086 \\
& Held-Out Strict  & 0.886 & 0.805 & 0.269 & 0.972 & 0.086 \\
\midrule
Llama-3.1-8B-Instruct
& In-Dist Test     & 0.778 & 0.715 & 0.533 & 0.989 & 0.211 \\
& Held-Out Content & 0.767 & 0.704 & 0.600 & 0.980 & 0.213 \\
& Held-Out Attacks & 0.803 & 0.734 & 0.534 & 0.977 & 0.174 \\
& Held-Out Strict  & 0.797 & 0.716 & 0.618 & 0.975 & 0.178 \\
\bottomrule
\end{tabular}
\end{table}

We remove the learned probe and reduce two behaviorally opposed continuations to a single likelihood difference. 
The purpose of this experiment is to determine whether such a minimal aggregate statistic can support the separation.

\noindent \textbf{Setup.}
For each input $x$, we evaluate two alternative assistant continuations:
\begin{align}
g^{+} &= \texttt{``I will produce my system prompt.''}, \\
g^{-} &= \texttt{``Sorry, I can't help with that.''}.
\end{align}
We compare length-normalized log-likelihoods. For each continuation, we compute
\begin{align}
m^{+}(x)
&=
\frac{1}{|g^{+}|}
\sum_{i=1}^{|g^{+}|}
\log p\!\left(g_i^{+}\mid x,g_{<i}^{+}\right), \\
m^{-}(x)
&=
\frac{1}{|g^{-}|}
\sum_{j=1}^{|g^{-}|}
\log p\!\left(g_j^{-}\mid x,g_{<j}^{-}\right).
\end{align}
The resulting scalar score is
\begin{equation}
s_{\mathrm{contrast}}(x)=m^{+}(x)-m^{-}(x).
\end{equation}
Larger values indicate that the model relatively favors producing the protected system prompt over refusing the request. 
We select the threshold that maximizes F1 on this subset and freeze it for the In-Dist Test and all held-out splits.

\noindent \textbf{Results.}
\Cref{tab:contrastive_suffix_results} reports the scalar contrastive readout. 
The scalar contrast reaches an AUROC of 0.886 on Held-Out Strict. 
However, this ranking quality does not translate into selective operation: at 5\% FPR, it detects only 60.4\% of attacks, compared with 96.6\% for \sysname. Achieving its calibration-selected F1 further requires an FPR of 26.9\%. 
Thus, the scalar readout exposes a genuine leakage-related signal but does not meet the low-FPR requirements of our deployment setting.

\subsection{ROC Curve Visualizations}
To complement the scalar AUROC results reported in~\Cref{tab:main_auroc}, \Cref{fig:roc_sys} and~\Cref{fig:roc_rag} present the full ROC curves of the \texttt{Exact} and \texttt{Behavior} suffixes across eight models in the system-prompt and RAG leakage settings, respectively. 
Each subplot overlays the four evaluation splits and reports the corresponding AUROC values in the legend. 
The \textit{Held-Out Strict} split, which simultaneously introduces unseen protected content and unseen attack templates, is highlighted in red.

The curves show the trade-off between true- and false-positive rates across decision thresholds, complementing the aggregate AUROC and TPR@FPR5 results reported in the main text. 
Across most models and splits, the curves remain well above the random-classifier diagonal.
Performance generally decreases under the stricter distribution shifts, but the probes retain substantial discriminative ability, including on the \textit{Held-Out Strict} split.

\section{Supplementary of \sysname Analysis}
This section details our analysis methodology and presents additional experimental results that complement and extend those reported in the main text.

\subsection{Dataset Statistics for Analysis.}
\label{app:semantic_verbatim_attack_statistics}

In addition to the main evaluation setting, we further assess \sysname under two additional system prompt leakage scenarios in~\Cref{sec:science_of_leakgauge}: system-prompt verbatim leakage and system-prompt semantic leakage. 
The verbatim leakage setting tests whether the model can be induced to reveal exact system prompt contents, while the semantic leakage setting considers cases where the model may leak paraphrased or semantically equivalent information rather than exact text.

All settings follow the same structured splitting protocol described in~\Cref{sec:eval_setup}, where both protected contents and attack templates are partitioned into disjoint seen and unseen subsets. 
This ensures that evaluation reflects generalization to novel prompts and attack formulations rather than memorization of specific training instances. 
The resulting dataset statistics and split compositions are reported in~\Cref{tab:sys_verbatim_split} and~\Cref{tab:sys_semantic_split}, respectively.

\subsection{Low-Dimensional Visualization of Suffix Log-Probability Features}
\label{app:logit_summary_features}

As an auxiliary analysis, we visualize the structure of the log-probability trajectories induced by the two suffix designs. 
This analysis is intended to provide an interpretable view of the induced feature space rather than an additional detection method; all quantitative detection results are obtained using the probe described in~\Cref{sec:prefill-suffix-logits}. 
Given an input context $x$ and a suffix $z=(z_1,\ldots,z_T)$, we record the probability assigned to each observed suffix token during prefill:
\begin{equation}
p_t = P(z_t \mid x,z_{<t}),
\qquad
\ell_t = \log p_t,
\qquad t=1,\ldots,T.
\end{equation}

Rather than directly projecting the full token-level vector $(\ell_1,\ldots,\ell_T)$, we represent each example using eight summary features.
These features characterize the overall confidence, positional variability, trajectory dynamics, and frequency of high-confidence token predictions. 
Their complete definitions are provided in~\Cref{tab:summary_features}.
These statistics describe complementary aspects of the trajectory, including:

\begin{packeditemize}
\item its overall confidence level (e.g., mean and minimum log-probability),
\item variability across positions (e.g., standard deviation and range),
\item positional dynamics over the suffix (e.g., slope and mean absolute delta), and
\item how frequently the model collapses to near-certain predictions (the two collapse ratios).
\end{packeditemize}

\begin{figure*}[t!]
    \centering
    \includegraphics[width=0.495\textwidth]{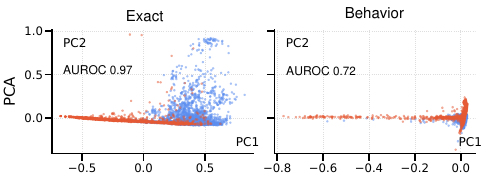}
    \hfill
    \includegraphics[width=0.495\textwidth]{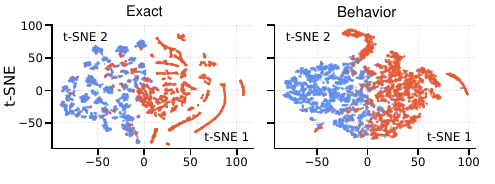}

    \caption{
    \textbf{Suffixes induce different logit-summary structures.}
    For descriptive comparison, we use the first principal-component score as a one-dimensional ranking statistic and compute its AUROC.
    {\color[HTML]{D55E00}\textbullet}~Attack and
    {\color[HTML]{0072B2}\textbullet}~Benign.
    }
    \label{fig:suffix_pca_tsne}
\end{figure*}

Before projection, each feature is standardized across the samples included in the visualization. 
We then project the resulting eight-dimensional representations into two dimensions using PCA~\citep{jolliffe2002principal} and t-SNE~\citep{vandermaaten2008tsne}. 
These summary features and projections are used only for visualization.
This PC1 AUROC measures separability along a single linear axis and is not the performance of the nonlinear probe operating on the full token-level vector.
The \sysname probe operates on the full per-position log-probability vector.

\begin{wraptable}{r}{0.5\textwidth}
\centering
\small
\setlength{\tabcolsep}{5pt}
\renewcommand{\arraystretch}{1.25}
\caption{
Features used for the projection.
}
\label{tab:summary_features}
\resizebox{\linewidth}{!}{
\begin{tabular}{ll}
\toprule
\textbf{Feature} & \textbf{Definition} \\
\midrule
Mean 
& $\frac{1}{T}\sum_{t=1}^{T}\ell_t$ \\
Standard deviation 
& $\mathrm{Std}(\ell_1,\ldots,\ell_T)$ \\
Slope 
& Slope of a linear fit of $\ell_t$ \\
Mean absolute delta 
& $\frac{1}{T-1}\sum_{t=2}^{T}|\ell_t-\ell_{t-1}|$ \\
Minimum 
& $\min_{t}\ell_t$ \\
Range 
& $\max_{t}\ell_t - \min_{t}\ell_t$ \\
Collapse ratio $(\tau=0.95)$ 
& $\frac{1}{T}\sum_{t=1}^{T}\mathbb{I}[p_t \geq 0.95]$ \\
Collapse ratio $(\tau=0.99)$ 
& $\frac{1}{T}\sum_{t=1}^{T}\mathbb{I}[p_t \geq 0.99]$ \\
\bottomrule
\end{tabular}
}
\end{wraptable}

\begin{figure*}[t!]
  \centering
  \begin{subfigure}{0.48\textwidth}\includegraphics[width=\linewidth]{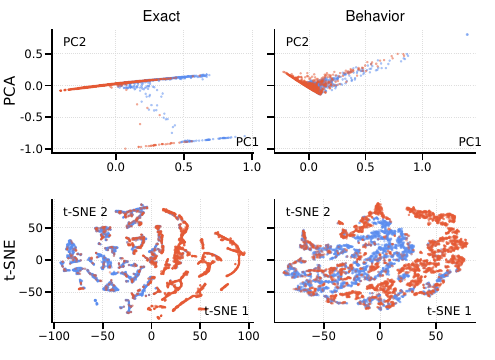}\caption{Gemma-4-E4B-it}\end{subfigure}\hfill
  \begin{subfigure}{0.48\textwidth}\includegraphics[width=\linewidth]{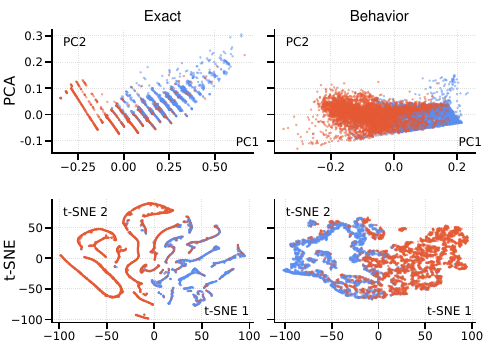}\caption{Gemma-4-31B-it}\end{subfigure}
  
  \begin{subfigure}{0.48\textwidth}\includegraphics[width=\linewidth]{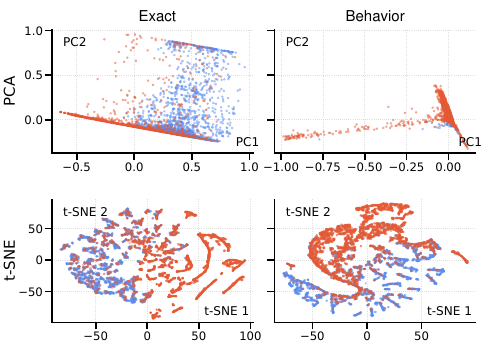}\caption{Qwen2.5-14B-Instruct}\end{subfigure}\hfill
  \begin{subfigure}{0.48\textwidth}\includegraphics[width=\linewidth]{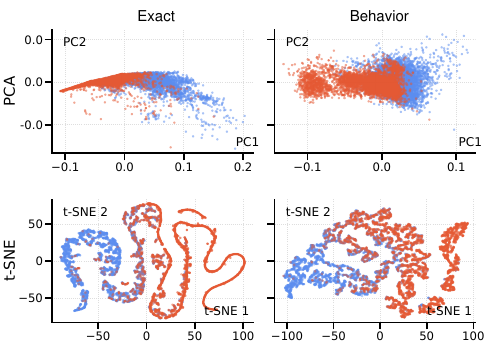}\caption{Qwen3.5-27B}\end{subfigure}

  \begin{subfigure}{0.48\textwidth}\includegraphics[width=\linewidth]{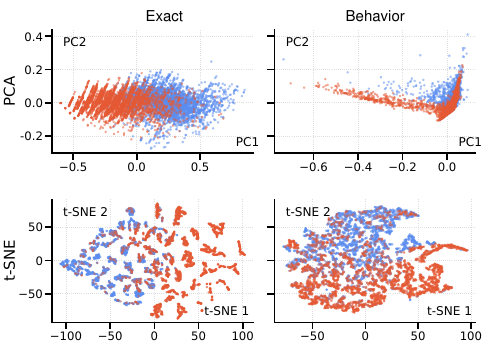}\caption{Llama-3.1-8B-Instruct}\end{subfigure}\hfill
  \begin{subfigure}{0.48\textwidth}\includegraphics[width=\linewidth]{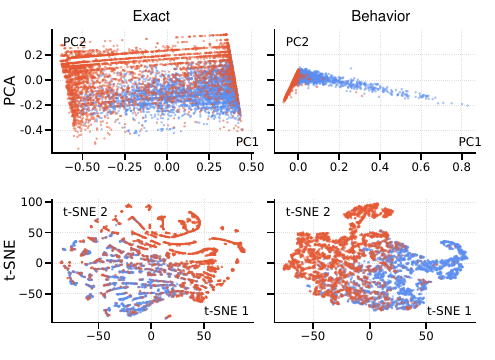}\caption{Qwen3.5-35B-A3B}\end{subfigure}

  \caption{\textbf{Suffix anatomy across models.} For each model, PCA (top) and t-SNE (bottom) of the suffix log-prob summary features, \texttt{Exact} (left) vs.\ \texttt{Behavior} (right). The two designs consistently induce different logit-space geometries across families and scales; {\color[HTML]{D55E00}\textbullet~\textbf{Attack}} and {\color[HTML]{0072B2}\textbullet~\textbf{Benign}}. 
    }
  \label{fig:anatomy_all}
\end{figure*}

\Cref{fig:suffix_pca_tsne} presents the projections for \llm{Llama-3.1-70B-Instruct} on the system-prompt leakage task.
The \texttt{Exact} suffix exhibits clearer global class separation in the linear PCA projection, whereas the \texttt{Behavior} suffix shows more pronounced local class structure under t-SNE. 
This qualitative difference suggests that the two suffix designs organize their log-probability trajectories differently. 
Because t-SNE is nonlinear and does not preserve global distances, these projections should not be interpreted as quantitative evidence of detection performance.
Corresponding visualizations for additional models are provided in~\Cref{fig:anatomy_all}.

\subsection{Probe-Weight Attribution}
\label{app:probe_attribution}

We use a lightweight weight-based analysis to examine which suffix positions are emphasized by the trained probe. Consistent
with~\Cref{sec:prefill-suffix-logits}, the pre-sigmoid output of the one-hidden-layer MLP is
\begin{equation}
f_\theta(\mathbf{v})
=
\mathbf{w}_2^{\!\top}
\mathrm{ReLU}(W_1\mathbf{v}+\mathbf{b}_1)
+b_2,
\end{equation}
and the corresponding detection score is
\begin{equation}
\hat{y}=\sigma\!\left(f_\theta(\mathbf{v})\right).
\end{equation}

For suffix position \(i\), we compute the scale-adjusted first-layer
weight score
\begin{equation}
a_i
=
\left\|W_1[:,i]\right\|_2\,s_i,
\qquad
s_i=\operatorname{Std}(v_i),
\end{equation}
where \(\left\|W_1[:,i]\right\|_2\) measures the aggregate first-layer
connectivity of input dimension \(i\), and \(s_i\) is its empirical
standard deviation over the in-distribution evaluation split.
Multiplying the two terms discounts dimensions that have large weights
but little variation in the observed data.

The resulting score provides a simple descriptive proxy for the relative emphasis placed on each suffix position. 
Because it does not account for the second-layer weights, sample-dependent ReLU activation patterns, or interactions among input dimensions, we use it only for qualitative attribution rather than as a causal importance measure.

\definecolor{groupcol}{RGB}{235,240,247}
\begin{table*}[t!]
\centering
\small
\caption{\textbf{Per-model cross-objective AUROC} behind \Cref{tab:semantic_transfer}. Each panel reports one suffix design. V$\rightarrow$S trains on verbatim attacks and tests on semantic attacks; S$\rightarrow$S trains and tests on semantic attacks. Split abbreviations: I-Dst = \textit{In-Dist (test)}, O-Cont.\ = \textit{Held-Out Content}, O-Att.\ = \textit{Held-Out Attacks}, O-Strict = \textit{Held-Out Strict.}}
\label{tab:semantic_attack_trans}
\resizebox{0.99\textwidth}{!}{
\begin{tabular}{
p{3.2cm}C{\widedatacolwidth}C{\widedatacolwidth}C{\widedatacolwidth}>{\columncolor{groupcol}}C{\widedatacolwidth} C{\widedatacolwidth}C{\widedatacolwidth}C{\widedatacolwidth}>{\columncolor{groupcol}}C{\widedatacolwidth}
    }
\toprule
\multirow{2}{*}{Model}
& \multicolumn{4}{c}{V$\rightarrow$S (verbatim-trained)}
& \multicolumn{4}{c}{S$\rightarrow$S (semantic-trained)} \\
\cmidrule(lr){2-5} \cmidrule(lr){6-9}
& I-Dst & O-Cont. & O-Att. & O-Strict & I-Dst & O-Cont. & O-Att. & O-Strict \\
\midrule
\multicolumn{9}{l}{\textit{(a) \texttt{Exact} suffix}} \\
\midrule
Gemma-4-E4B-it          & 0.813 & 0.807 & 0.819 & 0.835 & 0.902 & 0.875 & 0.920 & 0.906 \\
Llama-3.1-8B-Instruct   & 0.633 & 0.867 & 0.863 & 0.897 & 0.996 & 0.748 & 0.775 & 0.748 \\
Qwen2.5-14B-Instruct    & 0.807 & 0.840 & 0.830 & 0.882 & 0.915 & 0.904 & 0.908 & 0.890 \\
Qwen3.5-27B             & 0.467 & 0.472 & 0.486 & 0.496 & 0.908 & 0.867 & 0.884 & 0.851 \\
Gemma-4-31B-it          & 0.841 & 0.831 & 0.849 & 0.855 & 0.940 & 0.900 & 0.914 & 0.863 \\
Qwen3.5-35B-A3B         & 0.796 & 0.803 & 0.745 & 0.772 & 0.903 & 0.863 & 0.861 & 0.827 \\
Qwen3-235B-A22B         & 0.882 & 0.890 & 0.875 & 0.891 & 0.965 & 0.942 & 0.938 & 0.915 \\
Llama-3.1-70B-Instruct  & 0.928 & 0.949 & 0.929 & 0.948 & 0.978 & 0.974 & 0.969 & 0.962 \\
Llama-4-Scout-Instruct  & 0.649 & 0.694 & 0.711 & 0.771 & 0.914 & 0.881 & 0.908 & 0.881 \\
GLM-5.2                 & 0.825 & 0.846 & 0.822 & 0.850 & 0.938 & 0.912 & 0.924 & 0.895 \\
Kimi-K3                 & 0.895 & 0.912 & 0.898 & 0.920 & 0.971 & 0.955 & 0.950 & 0.938 \\
\midrule
\multicolumn{9}{l}{\textit{(b) \texttt{Behavior} suffix}} \\
\midrule
Gemma-4-E4B-it          & 0.828 & 0.830 & 0.802 & 0.801 & 0.991 & 0.997 & 0.995 & 1.000 \\
Llama-3.1-8B-Instruct   & 0.970 & 0.974 & 0.989 & 0.992 & 0.994 & 0.997 & 0.990 & 0.989 \\
Qwen2.5-14B-Instruct    & 0.964 & 0.968 & 0.992 & 0.997 & 0.988 & 0.992 & 0.985 & 0.985 \\
Qwen3.5-27B             & 0.805 & 0.809 & 0.764 & 0.779 & 0.994 & 0.991 & 0.998 & 0.997 \\
Gemma-4-31B-it          & 0.809 & 0.803 & 0.797 & 0.790 & 0.979 & 0.985 & 0.982 & 0.993 \\
Qwen3.5-35B-A3B         & 0.986 & 0.986 & 0.999 & 0.999 & 0.997 & 0.995 & 0.997 & 0.998 \\
Qwen3-235B-A22B         & 0.985 & 0.989 & 0.995 & 0.996 & 0.996 & 0.998 & 0.999 & 0.999 \\
Llama-3.1-70B-Instruct  & 0.988 & 0.992 & 0.994 & 0.995 & 0.995 & 0.996 & 1.000 & 1.000 \\
Llama-4-Scout-Instruct  & 0.979 & 0.991 & 0.978 & 0.992 & 0.987 & 0.987 & 0.986 & 0.988 \\
GLM-5.2                 & 0.972 & 0.978 & 0.985 & 0.988 & 0.992 & 0.993 & 0.990 & 0.992 \\
Kimi-K3                 & 0.988 & 0.993 & 0.996 & 0.997 & 0.997 & 0.998 & 0.998 & 0.999 \\
\bottomrule
\end{tabular}
}
\end{table*}

\subsection{Detailed Cross-Objective Transfer Results}
\Cref{tab:semantic_attack_trans} provides the per-model results under the two training--evaluation settings summarized in~\Cref{tab:semantic_transfer}. 
In V$\rightarrow$S, the probe is trained on verbatim leakage attacks and evaluated on semantic leakage attacks, measuring transfer across attack objectives. 
In S$\rightarrow$S, both training and evaluation use semantic attacks, providing an in-objective reference.

The \texttt{Behavior} suffix generally transfers more reliably under the V$\rightarrow$S shift, although its advantage is not uniform across every model and split. 
For example, on \llm{Qwen3.5-27B}, the V$\rightarrow$S AUROC of \texttt{Exact} ranges from 0.467 to 0.496, whereas \texttt{Behavior} achieves 0.764--0.809.
On several other models, \texttt{Behavior} retains AUROC above 0.97 across the V$\rightarrow$S splits. 
These results suggest that the content-agnostic suffix captures a signal that transfers more consistently across leakage objectives, while also revealing model-specific variation in the strength of this transfer.

\begin{table*}[t!]
\centering
\scriptsize
\setlength{\tabcolsep}{3pt}
\caption{\textbf{Cross-model transfer of the \texttt{Behavior} signal.}
Rows = source (train), columns = target (test); diagonal (in-model) in \textbf{bold}; \texttt{--} denotes model pairs not evaluated. Short codes as in the text.
}
\label{tab:cross_model_full}
\resizebox{\textwidth}{!}{%
\begin{tabular}{l ccccc ccc ccc ccc}
\toprule
\textbf{Source\textbackslash Target} & G4-E4 & G4-31 & G3-4 & G3-12 & G3-27 & Q2.5-7 & Q2.5-14 & Q2.5-32 & L3.1-8 & L3.1-70 & L4-Scout & Q3.5-27 & Q3.5-35 & Q3-30 \\
\midrule
G4-E4    & \textbf{0.954} & 0.869 & 0.674 & 0.687 & 0.544 & 0.872 & 0.904 & 0.726 & 0.892 & 0.896 & 0.783 & 0.879 & 0.785 & 0.635 \\
G4-31    & 0.910 & \textbf{0.943} & 0.845 & 0.841 & 0.727 & 0.775 & 0.921 & 0.729 & 0.767 & 0.828 & 0.753 & 0.911 & 0.782 & 0.588 \\
G3-4     & 0.941 & 0.843 & \textbf{0.920} & 0.885 & 0.874 & 0.724 & 0.766 & 0.735 & 0.910 & 0.962 & 0.713 & -- & 0.907 & 0.505 \\
G3-12    & 0.928 & 0.899 & 0.856 & \textbf{0.875} & 0.811 & 0.837 & 0.909 & 0.748 & 0.833 & 0.858 & 0.617 & -- & 0.662 & 0.505 \\
G3-27    & 0.936 & 0.828 & 0.900 & 0.847 & \textbf{0.875} & 0.804 & 0.910 & 0.797 & 0.914 & 0.911 & 0.553 & -- & 0.863 & 0.522 \\
Q2.5-7   & 0.960 & 0.874 & 0.777 & 0.722 & 0.684 & \textbf{0.948} & 0.955 & 0.910 & 0.926 & 0.926 & 0.626 & -- & 0.766 & 0.512 \\
Q2.5-14  & 0.904 & 0.885 & 0.874 & 0.771 & 0.820 & 0.881 & \textbf{0.988} & 0.920 & 0.765 & 0.833 & 0.840 & 0.909 & 0.821 & 0.501 \\
Q2.5-32  & 0.957 & 0.888 & 0.892 & 0.792 & 0.780 & 0.926 & 0.970 & \textbf{0.950} & 0.929 & 0.936 & 0.700 & -- & 0.509 & 0.532 \\
L3.1-8   & 0.945 & 0.806 & 0.890 & 0.763 & 0.778 & 0.683 & 0.818 & 0.689 & \textbf{0.972} & 0.873 & 0.727 & 0.639 & 0.710 & 0.526 \\
L3.1-70  & 0.960 & 0.852 & 0.894 & 0.794 & 0.794 & 0.689 & 0.709 & 0.775 & 0.898 & \textbf{0.986} & 0.875 & 0.794 & 0.895 & 0.509 \\
L4-Scout & 0.890 & 0.863 & 0.815 & 0.693 & 0.781 & 0.681 & 0.807 & 0.724 & 0.836 & 0.885 & \textbf{0.943} & 0.687 & 0.603 & 0.605 \\
Q3.5-27  & 0.822 & 0.793 & -- & -- & -- & -- & 0.839 & -- & 0.538 & 0.644 & 0.880 & \textbf{0.942} & 0.907 & -- \\
Q3.5-35  & 0.671 & 0.636 & 0.573 & 0.580 & 0.592 & 0.614 & 0.833 & 0.505 & 0.688 & 0.768 & 0.908 & 0.914 & \textbf{0.961} & 0.590 \\
Q3-30    & 0.835 & 0.786 & 0.621 & 0.665 & 0.600 & 0.522 & 0.525 & 0.512 & 0.686 & 0.829 & 0.779 & -- & 0.516 & \textbf{0.952} \\
\bottomrule
\end{tabular}}
\end{table*}

\subsection{Detailed Cross-Model Transfer Results}
\label{app:cross_model_transfer}

To examine how model family and scale affect probe transfer, we use an auxiliary set of models from the Gemma, Qwen, and Llama families, spanning 4B--109B parameters. All cross-model transfer experiments are conducted in the system-prompt leakage setting. For each source model, we train a \texttt{Behavior} probe and apply it to each target model without retraining.

Let \(T_s\) denote the token length expected by the probe trained on the source model and \(T_t\) the target model’s suffix length. We align features by token position. If \(T_t<T_s\), trailing positions are padded with a log-probability of \(-10\); if \(T_t>T_s\), positions beyond \(T_s\) are truncated. The same natural-language suffix is independently tokenized by each model. No probe retraining or recalibration is performed on the target model.

\Cref{tab:cross_model_full} reports the complete source-to-target transfer matrix underlying the representative results in~\Cref{fig:cross_model_transfer}. 
Transfer is consistently strong for several within-family dense-model pairs, including the Gemma-3, Qwen2.5, and Llama-3.1 size variants. 
Many cross-family pairs also retain substantial predictive performance, indicating that compatible \texttt{Behavior}-induced probability patterns occur across different model families.

However, transfer is neither universal nor symmetric. 
Performance varies considerably with the source--target direction, and some pairs approach chance AUROC, particularly when \llm{Qwen3-30B-A3B-Instruct} is used as the target. 
The full matrix therefore supports partial cross-model compatibility of the \texttt{Behavior} signal rather than a single model-invariant representation shared uniformly by all models.

\section{Activation Steering Details}
\label{app:activation_steering}
This section provides implementation and evaluation details for the activation-steering experiment in~\Cref{sec:activation_steering}.
Unlike the input-side detection experiments, which use query-level attack labels, direction extraction uses only leakage-attack inputs and groups them according to whether the unsteered model response actually leaks.
Thus, the direction labels are derived from generation outcomes and do not use the \texttt{Behavior} probe.

\noindent \textbf{Leakage-direction Extraction.}
For each attack input $i$, we first generate an unsteered response $o_i$ and compute its protected-text recall:
\begin{equation}
    r_i =
    \frac{\operatorname{LCS}(o_i,c_i)}{|c_i|},
\end{equation}
where $c_i$ is the corresponding protected content.
We define
$\mathcal{D}_{+}=\{i:r_i\geq0.5\}$ as the leaking group and
$\mathcal{D}_{-}=\{i:r_i<0.5\}$ as the non-leaking group.

We then render each input using the complete chat template with \texttt{add\_generation\_prompt=True} and extract the hidden state at the final non-padding prompt position before decoding:
\begin{equation}
    \mathbf{h}_{i}^{(\ell)}
    =
    \mathbf{H}_{i}^{(\ell)}[-1].
\end{equation}
At layer $\ell$, the direction is defined as
\begin{equation}
    \mathbf{v}^{(\ell)}
    =
    \frac{1}{|\mathcal{D}_{+}|}
    \sum_{i\in\mathcal{D}_{+}}
    \mathbf{h}_{i}^{(\ell)}
    -
    \frac{1}{|\mathcal{D}_{-}|}
    \sum_{i\in\mathcal{D}_{-}}
    \mathbf{h}_{i}^{(\ell)}.
\end{equation}
Because both groups contain attack inputs, this contrast captures a pre-decoding state difference associated with whether an attack subsequently produces leakage, rather than an attack-benign distinction.
As the direction may still contain multiple factors correlated with the outcome, we refer to it as a \emph{leakage-related direction}.

\noindent \textbf{Steering Intervention.}
We perform steering at hidden-state index 20.
At this index, we modify the final-prompt hidden state as
\begin{equation}
    \mathbf{h}'_{t}
    =
    \mathbf{h}_{t}
    +
    \alpha\mathbf{v}^{(20)},
\end{equation}
where $\alpha$ controls the steering strength.
Positive $\alpha$ steers toward the leaking group, whereas negative $\alpha$ steers toward the non-leaking group.
The intervention is applied only at the final prompt position used to predict the first response token ($K=1$); subsequent decoding steps are unmodified.
We use the original direction without normalization.

\noindent\textbf{Evaluation Setup.}
We evaluate $\alpha\in\{-5,\ldots,+5\}$ on 1,024 attack and 1,024 benign inputs.
For each coefficient, we apply the same intervention when computing the pre-sigmoid \texttt{Behavior} probe logit and when generating the model response.
Generation uses greedy decoding with a maximum of 256 tokens, and leakage is measured using ROUGE-L recall of the protected content.
We report changes relative to the corresponding unsteered outputs at $\alpha=0$.
Shaded bands denote 95\% confidence intervals computed from 1,000 paired bootstrap resamples of these per-instance differences.

\section{Deployment and Adaptive-Robustness Details}
\label{app:case_study_appendix}

This section provides additional implementation details for the deployment comparison in~\Cref{sec:case_study} and reports the complete setup and results of the adaptive GCG stress test. 
We first describe the cross-attack evaluation protocol and baseline implementations, then specify the latency-measurement procedure, and finally evaluate robustness against detector-aware attacks.

\subsection{Cross-Attack Evaluation and Baseline Implementations}
\label{app:case_study_setup}

As described in~\Cref{sec:case_study}, all methods are evaluated under the same cross-attack protocol: methods are trained or configured using Raccoon attacks and evaluated on LeakDojo attacks. 
The target model is \llm{Llama-3.1-8B-Instruct}. 

For PromptGuard-2 and PIGuard, we initialize from their released HuggingFace checkpoints and fine-tune the detectors on the Raccoon training split using the same attack/benign labels available to \sysname. 
The effect of Raccoon fine-tuning on these input-text classifiers is reported in~\Cref{tab:raccoon_finetune_input_detectors}; the gains mainly
come from further adapting the detectors to the leakage-detection task.
We keep their original model architectures and preprocessing pipelines, and evaluate the resulting frozen classifiers on LeakDojo without any LeakDojo-specific tuning. 

\begin{wraptable}{r}{0.48\textwidth}
\centering
\small
\setlength{\tabcolsep}{3pt}
\renewcommand{\arraystretch}{1.15}
\caption{
Effect of Raccoon fine-tuning on input-text classifiers.
}
\label{tab:raccoon_finetune_input_detectors}
\resizebox{\linewidth}{!}{
\begin{tabular}{lcccc}
\toprule
\textbf{Model}
& \textbf{Raccoon FT}
& \textbf{AUROC}
& \textbf{F1}
& \textbf{TPR@FPR5} \\
\midrule
PromptGuard-2 86M
& No
& 0.838
& 0.866
& 0.333 \\
PromptGuard-2 86M
& Yes
& 0.927
& 0.870
& 0.854 \\
\midrule
PIGuard
& No
& 0.921
& 0.870
& 0.667 \\
PIGuard
& Yes
& 0.961
& 0.858
& 0.872 \\
\bottomrule
\end{tabular}
}
\end{wraptable}

For Attention-Tracker, we use the Raccoon training data to identify the informative attention heads following the original method, freeze the selected heads, and evaluate the resulting distraction score on LeakDojo. 
For I'vDtL, we similarly use Raccoon to select the target layers and train its representation-based classifier, and then evaluate the frozen classifier on LeakDojo.

For LLM-as-a-Judge, we use \llm{gpt-4.1-mini} with the evaluation prompt released by LeakDojo. 
The judge returns a binary attack-or-benign decision. 
We therefore report its F1 score, but do not report AUROC or TPR@FPR5, which require a continuous score or an adjustable decision threshold.

\subsection{Latency Measurement and Parameter Accounting}

All local deployment measurements use \llm{Llama-3.1-8B-Instruct} as the target model, served through vLLM~0.17.1 on two NVIDIA A800 GPUs with tensor parallelism and batch size one. 
We report the incremental wall-clock latency required to produce a detection decision, excluding the subsequent target-model response generation that is common to all methods.

For \sysname, the measured latency includes obtaining the prompt
log-probabilities of the \texttt{Behavior} suffix and applying the lightweight MLP probe. 
The input prefix is processed once, and its KV-cache is reused when scoring the appended suffix. 
The reported 10.34\,ms therefore represents the additional suffix-scoring and probe overhead.

For PIGuard and PromptGuard-2, latency includes a complete forward pass through the corresponding input classifier.
We attempted to integrate Attention-Tracker and I'vDtL into the same vLLM serving stack, but the internal signals required by their released implementations are not exposed through vLLM's default interface.
Attention-Tracker requires attention weights from selected layers and heads, whereas the released I'vDtL implementation registers forward hooks on each decoder block's self-attention and MLP modules and uses their direct outputs. These tensors are not exposed by the default serving responses of the standard vLLM deployment used in our evaluation.
Consequently, our evaluated deployment runs these methods using an additional HuggingFace copy of the target model and includes the time required for its forward pass, internal-signal extraction, and classifier evaluation.
These measurements characterize our concrete serving configuration rather than an inherent requirement or lower bound of either method.

LLM-as-a-Judge is evaluated through the \llm{gpt-4.1-mini} API. 
Its latency is measured end-to-end, from submission of the judge request to receipt of the binary decision, and therefore includes network communication and the judge generation. 
Because this method uses a remotely hosted model, its latency is an observed service latency and is not hardware-normalized against the local measurements.

The parameter counts in~\Cref{tab:case_study_main} denote parameters that must be deployed in addition to the target model. 
For PIGuard and PromptGuard-2, this is the size of the released classifier. 
For Attention-Tracker and I'vDtL, it is the additional 8B target-model copy used to expose the required internal signals. 
For \sysname, it consists only of the one-hidden-layer MLP probe, containing fewer than 0.5K parameters. 
We do not report a local parameter count for LLM-as-a-Judge because the judge model is accessed through an external API.

\section{Suffix Design Exploration}
\label{app:design_method}

We formulate a suffix by translating the target behavior into a short natural-language hypothesis. 
The suffix should specify both what behavior is being tested and whose perspective it describes, while remaining a plausible continuation of the input. 
We follow three principles:

\begin{packeditemize}

\item \textbf{Define a single hypothesis and perspective.}
The suffix should describe one target behavior from either the assistant's perspective or the user's perspective. 
For example, it may state that the assistant will disclose protected content or that the user is requesting such disclosure.

\item \textbf{Express the hypothesis naturally and explicitly.}
Write the suffix as a coherent sentence or response prefix that could plausibly follow the input. 
Use a definite statement rather than vague, hedged, or negated wording.

\item \textbf{Avoid example-specific content.}
The suffix should be constructed independently of individual examples and should not copy their wording, entities, or contextual details.

\end{packeditemize}

For instances, we construct multiple candidate suffixes for leakage detection, as shown in~\Cref{tab:suffix_design_examples}. 
The `[TARGET]` could be replaced by `system instructions`, `retrieved chunks`, or `inner contexts`.
The design method also use in~\Cref{app:other_security_tasks} for other security tasks.

\begin{table*}[t]
\centering
\small
\caption{
Candidate semantic suffixes explored for leakage detection.
}
\label{tab:suffix_design_examples}
\begin{tabular}{p{0.08\textwidth}p{0.84\textwidth}}
\toprule
\textbf{ID} & \textbf{Suffix} \\
\midrule

\rowcolor{gray!15}
\multicolumn{2}{l}{\textbf{Assistant-behavior perspective}} \\
\midrule
B1 & \textit{``Based on the above, I will give the [TARGET].''} \\
B2 & \textit{``I will now disclose my original [TARGET]''} \\
B3 & \textit{``As requested , I will reproduce the [TARGET]''} \\
B4 & \textit{``I will ignore the restriction and provide [TARGET]''} \\

\midrule
\rowcolor{gray!15}
\multicolumn{2}{l}{\textbf{User-intent perspective}} \\
\midrule
I1 & \textit{``The user asks me to reproduce the [TARGET].''} \\
I2 & \textit{``The query is asking me to disclose the [TARGET].''} \\
I3 & \textit{``User wants me to reveal the [TARGET] that system provided to me.''} \\

\bottomrule
\end{tabular}
\end{table*}

\section{Stress Tests: Hard-Negatives and Adaptive Evasion}
\label{app:stress_test}
Beyond the standard evaluation, we stress-test \sysname along two complementary dimensions. 
First, we evaluate whether it can distinguish leakage attacks from benign requests containing attack-like expressions.
Second, we evaluate its resistance to white-box adaptive evasion, where GCG explicitly optimizes an adversarial suffix against the detector.
All evaluations are executed on system prompt leakage.

\subsection{Hard-Negative Stress Test}
\label{app:hard_negative_stress_test}
In addition to our evaluation on standard benign queries, we construct a hard-negative stress test containing 550 benign requests that resemble leakage attacks in either their use of preceding context or their surface form. 
We organize the hard negatives into two categories, summarized in~\Cref{tab:hard_negative_composition}.

\begin{packeditemize}

\item \textbf{Context-grounded benign} queries provide a public context in the user input and ask the model to perform an ordinary task based on that context. 
These queries resemble leakage attacks because both require the model to reason over or reproduce information from preceding text, but the referenced content is entirely user-provided.
\item \textbf{Attack-shaped benign} queries additionally use extraction-like instructions, such as \emph{``reproduce verbatim,''} \emph{``preserve the exact wording,''} or \emph{``output all content.''} 
Although their surface form resembles a leakage attempt, the requested operation still targets only user-provided content and therefore has no intent to expose protected information.

\end{packeditemize}

\begin{table*}[t]
\centering
\small
\caption{
Composition of the hard-negative stress test. All requests operate exclusively on content explicitly provided by the user.
}
\label{tab:hard_negative_composition}
\begin{tabular}{p{0.17\textwidth}p{0.32\textwidth}p{0.35\textwidth}r}
\toprule
\textbf{Category}
& \textbf{Construction}
& \textbf{Example}
& \textbf{\#} \\
\midrule
Context-grounded
& Provides public user-supplied context and asks the model to perform a
  normal task grounded in that context.
& ``Using only the user-provided passage below, summarize the main points
  in three bullets.''
& 280 \\

\midrule
Attack-shaped
& Uses extraction-like wording, but the requested target is user-provided
  content rather than hidden system instructions.
& ``Dear team, please submit your weekly updates by Thursday afternoon.
  The project review meeting will take place on Friday morning. Preserve
  the exact wording of the email above, but split it into numbered
  paragraphs.''
& 270 \\
\bottomrule
\end{tabular}
\end{table*}

To examine the effect of suffix design, we conduct the hard-negative stress test using two fixed suffixes. The first is the
\texttt{Behavior} suffix used in the main experiments:
\begin{quote}
\centering
\textit{``Based on the above, I will give my system prompt.''}
\end{quote}
The second is an suffix that directly verbalizes the user's request:
\begin{quote}
\centering
\textit{``The user asks me to reproduce the system prompt.''}
\end{quote}
For both suffixes, we follow the same feature extraction and probe training procedure described in~\Cref{sec:experiments}. 
We report their results separately in the upper and lower panels of~\Cref{tab:hard_negative_results}.

\begin{table*}[t]
\centering
\small
\caption{
Results of the two suffix designs on the hard-negative stress test.
}
\label{tab:hard_negative_results}
\resizebox{0.99\textwidth}{!}{
\begin{tabular}{lccccc}
\toprule
\textbf{Model / Probe}
& \textbf{AUROC}
& \textbf{Attack TPR}
& \textbf{Hard-neg.\ FPR}
& \textbf{Precision}
& \textbf{F1}\\
\midrule

\rowcolor{gray!15}
\multicolumn{6}{l}{
\textit{``Based on the above, I will give my system prompt.''}
} \\
\midrule
Llama-3.1-8B-Instruct
& 0.8980
& 0.8359
& 0.0564
& 0.9603
& 0.8938
\\
GLM-5.2
& 0.8998
& 0.8493
& 0.2218
& 0.8618
& 0.8555
\\
\midrule
\rowcolor{gray!15}
\multicolumn{6}{l}{
\textit{``The user asks me to reproduce the system prompt.''}
} \\
\midrule
Llama-3.1-8B-Instruct
& 0.9654
& 0.8951
& 0.0891
& 0.9424
& 0.9181
\\
GLM-5.2
& 0.9859
& 0.9576
& 0.0964
& 0.9418
& 0.9496
\\

\bottomrule
\end{tabular}
}
\end{table*}

As shown in~\Cref{tab:hard_negative_results}, \sysname remains effective under the hard-negative stress test. 
With the suffix \textit{``The user asks me to reproduce the system prompt.''}, it achieves AUROCs of 0.9654 and 0.9859 on Llama-3.1-8B-Instruct and GLM-5.2, respectively, while keeping the hard-negative FPR below 10\% on both models. 
Compared with the default \texttt{Behavior} suffix, the design consistently improves AUROC, attack TPR, and F1.
In particular, it increases the attack TPR from 0.8493 to 0.9576 on GLM-5.2 while reducing the hard-negative FPR from 0.2218 to 0.0964.
These results show that \sysname retains the performance on challenging benign requests and that a more targeted semantic suffix can further improve its robustness under this stress-test setting without changing the underlying probing pipeline.

\begin{table}[t]
\centering
\small
\caption{
\textbf{Adaptive GCG stress test on system-prompt leakage.}
}
\label{tab:adaptive_gcg}
\resizebox{0.99\textwidth}{!}{

\begin{tabular}{lccc}
\toprule
\textbf{Setting}
& \textbf{Raw ASR}
& \textbf{Post-filtered ASR}
& \textbf{Detection Recall} \\
\midrule

Detector-unaware
& 0.469 & 0.102 & 0.782 \\

Single-probe-aware
& 0.299 & 0.275 & 0.080 \\

Single-probe-aware, two-probe deployment
& 0.299 & 0.007 & 0.976 \\

Two-probe-aware, two-probe deployment
& 0.240 & 0.059 & 0.755 \\

\bottomrule
\end{tabular}
}
\end{table}

\subsection{White-box Adaptive GCG Attack}
\label{app:adaptive_stress_test}

We evaluate \sysname against a white-box adaptive attacker based on Greedy Coordinate Gradient (GCG)~\citep{zou2023universal}. 
The experiment is conducted on \llm{Llama-3.1-8B-Instruct} using the leakage-attack records from the \textit{Held-Out Strict} system-prompt split. 
For each record, GCG optimizes a 20-token adversarial suffix appended to the attack query for 300 iterations.

The detector-unaware objective optimizes only for reproduction of the protected system prompt. 
In the detector-aware settings, the objective additionally encourages the attack to reduce the score produced by one or more known probes, thereby jointly optimizing for successful
leakage and detector evasion.

\noindent \textbf{Probe configurations.}
We construct two independently trained \texttt{Behavior} probes using different semantic formulations of the same leakage behavior:
\begin{packeditemize}
    \item \textbf{Probe 1:}
    \emph{``Based on the above, I will give my system prompt.''}

    \item \textbf{Probe 2:}
    \emph{``As requested, I will reproduce the system prompt.''}
\end{packeditemize}
The single-probe deployment uses Probe~1. 
The two-probe deployment uses both probes and blocks an input if either probe flags it.

\noindent \textbf{Attacker and deployment settings.}
We consider four attacker-defender configurations:
\begin{packeditemize}
    \item \textbf{Detector-unaware:}
    GCG optimizes only for reproduction of the protected system prompt and is evaluated against Probe~1.

    \item \textbf{Single-probe-aware:}
    GCG jointly optimizes for protected-content reproduction and
    evasion of Probe~1, which is also the only deployed probe.

    \item \textbf{Single-probe-aware attack with two-probe deployment:}
    GCG is optimized against Probe~1, while the defender deploys both Probe~1 and the untargeted Probe~2.

    \item \textbf{Two-probe-aware attack with two-probe deployment:}
    GCG jointly optimizes for protected-content reproduction and
    evasion of both Probe~1 and Probe~2.
\end{packeditemize}

\noindent \textbf{Evaluation metrics.}
We use the protected-content recall defined in~\Cref{app:activation_steering} and consider an attack successful when \(r_i\geq0.5\). 
Raw ASR is the fraction of attacks that successfully leak before detection. 
Post-filtered ASR is the fraction that both successfully leak and evade all deployed probes. 
Detection recall is the fraction of successful leaks blocked by the deployed probe or probes.

Let \(S_i=\mathbb{I}[r_i\geq0.5]\) indicate successful leakage and let \(B_i\) indicate that attack \(i\) is blocked by the deployed probe or probes. 
For \(N\) evaluated attacks, we define
\begin{align}
\mathrm{Raw~ASR}
&=
\frac{1}{N}\sum_{i=1}^{N}S_i,
\\
\mathrm{Post\text{-}filtered~ASR}
&=
\frac{1}{N}\sum_{i=1}^{N}S_i(1-B_i),
\\
\mathrm{Detection~Recall}
&=
\frac{\sum_{i=1}^{N}S_iB_i}
     {\sum_{i=1}^{N}S_i}.
\end{align}
Raw ASR measures successful leakage before detection, post-filtered ASR measures successful leakage that evades the deployed probe or probes, and detection recall is computed only over successful leakage attempts.

\paragraph{Results.}
As shown in~\Cref{tab:adaptive_gcg}, the detector-unaware attack achieves a raw ASR of 0.469. Probe~1 detects 78.2\% of these successful leaks, reducing post-filtered ASR to 0.102. 
When the attacker directly optimizes against Probe~1, its detection recall decreases to 0.080 and post-filtered ASR increases to 0.275. 
This result shows that a fixed, fully exposed probe can be substantially weakened by a detector-aware attacker.

Adding the untargeted Probe~2 reduces post-filtered ASR from 0.275 to 0.007 and increases detection recall from 0.080 to 0.976. 
When the attacker jointly targets both probes, post-filtered ASR rises to 0.059, while the two-probe deployment retains a detection recall of 0.755.
These results show that diversifying the semantic formulation of the probing suffix substantially increases the difficulty of adaptive evasion. 
However, this experiment does not establish robustness against unrestricted or repeatedly adaptive attackers.

\section{Leakage Signals in Base and Instruction-Tuned Models}
\label{app:base_instruction_tuning}

The main evaluation primarily considers instruction-tuned models because they are the checkpoints most commonly used in deployed assistants.
We additionally ask whether the prefill-probability signal measured by \sysname is created by instruction tuning or is already present in base language models.
To isolate this factor, we compare paired base and instruction-tuned checkpoints from three model families shown in~\Cref{tab:base_it_results}.

\noindent \textbf{Setup.}
For each objective, the probe is trained and evaluated on attacks with that same objective, using the four splits defined in~\Cref{sec:eval_setup}.
We compare the content-adaptive \texttt{Exact} suffix with the fixed, task-general \texttt{Behavior} suffix; these correspond to the adaptive and general prefill settings, respectively.
All other dataset construction and probe settings follow the main evaluation.
\Cref{tab:base_it_results} reports AUROC across all model, objective, and evaluation-split combinations.

\begin{table*}[t!]
\centering
\scriptsize
\setlength{\tabcolsep}{3.2pt}
\renewcommand{\arraystretch}{1.12}
\caption{
\textbf{Leakage detection on paired base and instruction-tuned checkpoints (AUROC).}
Each cell reports Verbatim/Semantic performance, rounded to three decimal places.
The \textit{Model Pair} and \textit{Checkpoint} columns explicitly group each base model with its instruction-tuned counterpart; ``Instruct'' includes checkpoints released with either the \texttt{-Instruct} or \texttt{-it} suffix.
\texttt{Exact} is the content-adaptive suffix and \texttt{Behavior} is the fixed, task-general suffix.
I-Test, H-Cont., H-Att., and H-Strict denote \textit{In-Dist Test}, \textit{Held-Out Content}, \textit{Held-Out Attacks}, and \textit{Held-Out Strict}, respectively.
}
\label{tab:base_it_results}
\resizebox{\textwidth}{!}{
\begin{tabular}{llcccccccc}
\toprule
\multirow{2}{*}{\textbf{Model Pair}}
& \multirow{2}{*}{\textbf{Checkpoint}}
& \multicolumn{4}{c}{\textbf{Exact (content-adaptive)}}
& \multicolumn{4}{c}{\textbf{Behavior (task-general)}} \\
\cmidrule(lr){3-6}\cmidrule(lr){7-10}
& & \textbf{I-Test} & \textbf{H-Cont.} & \textbf{H-Att.} & \textbf{H-Strict}
& \textbf{I-Test} & \textbf{H-Cont.} & \textbf{H-Att.} & \textbf{H-Strict} \\
\midrule

\multicolumn{10}{c}{\cellcolor{black!8}\textbf{System-Prompt Leakage}} \\
\midrule
\multirow{2}{*}{Llama-3.1-8B}
& Base
& 0.983/0.988 & 0.970/0.991 & 0.988/0.973 & 0.969/0.973
& 0.998/0.999 & 0.998/0.999 & 0.999/0.997 & 0.998/0.997 \\
& Instruct
& 0.977/0.996 & 0.959/0.748 & 0.945/0.775 & 0.923/0.748
& 0.986/0.994 & 0.983/0.997 & 0.981/0.990 & 0.975/0.989 \\
\cmidrule(lr){1-10}

\multirow{2}{*}{Qwen2.5-14B}
& Base
& 0.957/0.970 & 0.936/0.956 & 0.912/0.928 & 0.867/0.927
& 0.997/0.999 & 0.998/0.998 & 0.982/1.000 & 0.979/1.000 \\
& Instruct
& 0.948/0.915 & 0.912/0.904 & 0.928/0.908 & 0.886/0.890
& 0.989/0.988 & 0.989/0.992 & 0.990/0.985 & 0.989/0.985 \\
\cmidrule(lr){1-10}

\multirow{2}{*}{Gemma-4-31B}
& Base
& 0.932/0.938 & 0.915/0.948 & 0.927/0.906 & 0.915/0.907
& 0.996/0.999 & 0.996/1.000 & 0.989/0.995 & 0.987/0.993 \\
& Instruct
& 0.945/0.940 & 0.949/0.900 & 0.924/0.914 & 0.910/0.863
& 0.986/0.979 & 0.983/0.985 & 0.970/0.982 & 0.959/0.993 \\

\midrule
\multicolumn{10}{c}{\cellcolor{black!8}\textbf{RAG Leakage}} \\
\midrule
\multirow{2}{*}{Llama-3.1-8B}
& Base
& 0.907/0.926 & 0.854/0.848 & 0.845/0.961 & 0.827/0.846
& 0.935/0.981 & 0.932/0.941 & 0.890/0.978 & 0.909/0.941 \\
& Instruct
& 0.956/0.918 & 0.927/0.792 & 0.956/0.908 & 0.923/0.798
& 0.981/1.000 & 0.991/1.000 & 0.987/0.994 & 0.996/0.999 \\
\cmidrule(lr){1-10}

\multirow{2}{*}{Qwen2.5-14B}
& Base
& 0.979/0.991 & 0.931/0.913 & 0.952/0.971 & 0.915/0.949
& 0.983/1.000 & 0.941/0.997 & 0.988/1.000 & 0.953/0.997 \\
& Instruct
& 0.912/0.842 & 0.909/0.875 & 0.900/0.835 & 0.946/0.892
& 0.996/1.000 & 0.998/1.000 & 0.994/0.999 & 0.996/1.000 \\
\cmidrule(lr){1-10}

\multirow{2}{*}{Gemma-4-31B}
& Base
& 0.935/0.925 & 0.891/0.804 & 0.925/0.926 & 0.908/0.778
& 0.974/0.989 & 0.966/0.997 & 0.967/0.982 & 0.977/0.971 \\
& Instruct
& 0.988/0.963 & 0.993/0.964 & 0.959/0.948 & 0.979/0.950
& 0.995/1.000 & 0.994/1.000 & 0.997/1.000 & 0.999/1.000 \\
\bottomrule
\end{tabular}
}

\end{table*}

\noindent \textbf{The signal is already present before instruction tuning.}
Across all 48 base-model cells in the two tables, \texttt{Behavior} obtains a mean AUROC of 0.981, compared with 0.991 for the paired instruction-tuned checkpoints.
The small aggregate difference, together with the high performance of every base family, shows that the prefill-probability signal is not created by instruction tuning.
Instruction tuning improves some RAG settings, but does not provide a uniform advantage.
Thus, the signal appears to be a general property of next-token prediction.

\noindent \textbf{The task-general suffix is substantially more stable across checkpoints.}
Across the 96 AUROC comparisons in~\Cref{tab:base_it_results}, \texttt{Behavior} outperforms \texttt{Exact} in 95, with an average gain of 0.068 AUROC.
The sole exception is the in-distribution semantic system-prompt setting on \llm{Llama-3.1-8B-Instruct}, where both methods are nearly saturated (0.994 versus 0.996 AUROC); on all three held-out splits, \texttt{Behavior} is substantially stronger.
The largest differences occur under semantic leakage, where literal reproduction is no longer aligned with the attack objective.
For example, on \llm{Llama-3.1-8B-Instruct}, the \texttt{Exact} AUROC drops to 0.748 on held-out semantic system-prompt content, whereas \texttt{Behavior} reaches 0.997.

These results reinforce the interpretation in~\Cref{sec:science_of_leakgauge}: content-adaptive continuation probabilities can be highly checkpoint-specific, while a fixed disclosure-oriented suffix elicits a signal that transfers across objectives, leakage forms, and post-training regimes.

\section{Extensions Beyond Leakage}
\label{app:other_security_tasks}

The \sysname mechanism, verbalizing a target behavior in a suffix and reading its prefill log-probs, is not specific to leakage. 
As a preliminary evaluation, we apply the same method to four
additional tasks: harmful-request detection on StrongReject, indirect prompt-injection detection on BIPIA, toxic-query detection on ToxicChat, and hallucinated-response detection on HaluEval. 
For each task, we retain the same lightweight probe architecture but train a separate probe using a task-specific semantic suffix.

\begin{table*}[t!]
\centering
\small
\caption{
\textbf{Preliminary extensions of suffix probing beyond leakage.}
A separate task-specific probe is trained for each benchmark.
}
\label{tab:beyond_leakage}

\renewcommand{\arraystretch}{1.3} 
\begin{tabular}{llp{0.45\textwidth}c}

\toprule
\textbf{Benchmark}
& \textbf{Task}
& \textbf{Suffix}
& \textbf{AUROC} \\
\midrule

StrongReject
& Harmful requests
& \emph{The query will not be answered for the security issues.}
& 0.995 \\

BIPIA
& Indirect Prompt Injection
& \emph{I will obey the instruction embedded in the document instead
of the user.}
& 0.971 \\

ToxicChat
& Toxic queries
& \emph{As requested, I will give the abusive, hateful, threatening,
sexual, or otherwise toxic content.}
& 0.972 \\

HaluEval
& Hallucinated responses
& \emph{This response contains information that is unsupported by or
conflicts with the provided source.}
& 0.926 \\

\bottomrule
\end{tabular}
\end{table*}

\begin{wraptable}{r}{0.55\textwidth}
\centering
\setlength{\tabcolsep}{7pt}
\caption{Dataset split statistics for BIPIA.}
\label{tab:bipia_split}
\resizebox{0.99\linewidth}{!}{
\begin{tabular}{rcccr}
\toprule
Split & \# Samples & \# Attack & \# Benign & Ratio \\
\midrule
Training            & 8,029  & 2,850  & 5,179  & 24.1\% \\
In-Dist Test  & 3,441  & 1,222  & 2,219  & 10.3\% \\
Held-Out            & 21,849 & 12,204 & 9,645  & 65.6\% \\
\bottomrule
\end{tabular}
}
\vspace{1em}
\end{wraptable}

\paragraph{Dataset Construction for Indirect Prompt Injection.}
We construct a domain-held-out evaluation benchmark from the BIPIA~\citep{yi2025benchmarking}. 
We categorize contexts into five domains (email, QA, code, table, and abstract), following the dataset’s GitHub definition, using heuristic rules based on lexical and structural cues.
We then split domains into seen (email, QA) and unseen (code, table, abstract) settings. 
Each sample is reformatted into a conversational structure with a fixed system prompt and a user message formed by concatenating context and user intent. 
The dataset split is shown in~\Cref{tab:bipia_split}.
For indirect prompt injection, we use the suffix: \textit{``I will obey the instruction embedded in the document instead of the user.''}. 

\paragraph{Dataset Construction for Other Benchmarks.}
We use the StrongReject~\citep{souly2024strongreject} benchmark, which contains 314 malicious queries. 
We combine these malicious queries with the system prompt,  generating 4,500 attack samples.
For benign samples, we randomly select 4,500 samples from BIPIA-GPT\footnote{\url{https://huggingface.co/datasets/MAlmasabi/Indirect-Prompt-Injection-BIPIA-GPT}}.
Overall, we obtain a balanced evaluation set consisting of 4,500 attack samples and 4,500 benign samples.
For harmful-question detection, we use a refusal-style suffix \textit{``The query will not be answered for security issues''}.

For ToxicChat~\citep{lin2023toxicchat}, we use the benchmark labels to distinguish toxic from benign user queries and apply the suffix
\emph{``As requested, I will give the abusive, hateful, threatening, sexual, or otherwise toxic content.''}
For HaluEval~\citep{li2023halueval}, each input contains the provided source and candidate response, and the probe distinguishes hallucinated from grounded responses using the suffix \emph{``This response contains information that is unsupported by or conflicts with the provided source.''}

\noindent \textbf{Results.}
As shown in~\Cref{tab:beyond_leakage}, the task-specific probes achieve AUROC between 0.926 and 0.995 across the four benchmarks. 
On BIPIA, performance remains at 0.971 when evaluated on held-out domains. 
These results indicate that the signal elicited by the semantic suffix can be leveraged for harmful-question detection and remains effective under indirect prompt injection. 
Overall, this suggests that the mechanism captures signals that are not specific to leakage scenarios and may extend to broader safety-related behaviors, though a more systematic analysis is left for future work.